\documentclass[reqno,12pt]{article}
\usepackage{amsmath,amsfonts,amssymb,amsthm,amstext,amscd,eucal,xcolor}
\usepackage[all]{xy}
\usepackage{hyperref}
\usepackage{epsfig}
\usepackage{color}
\usepackage{graphicx}
\usepackage[active]{srcltx}
\makeatletter \@addtoreset{equation}{section}

\makeatletter\renewcommand\section{\@startsection {section}{1}{\z@}%
                                   {-3.5ex \@plus -1ex \@minus -.2ex}%nn
                                   {2.3ex \@plus.2ex}%
                                   {\normalfont\large\bfseries}}
\renewcommand\subsection{\@startsection{subsection}{2}{\z@}%
                                     {-3.25ex\@plus -1ex \@minus -.2ex}%
                                     {1.5ex \@plus .2ex}%
                                     {\normalfont\bfseries}}

\newcommand{\be}{\begin{equation}}
\newcommand{\ee}{\end{equation}}
\newcommand{\bea}{\begin{eqnarray}}
\newcommand{\eea}{\end{eqnarray}}
\newcommand{\bse}{\begin{subequations}}
\newcommand{\ese}{\end{subequations}}
\newcommand{\bi}{\begin{itemize}}
\newcommand{\ei}{\end{itemize}}

\newcommand{\beq}{\begin{eqnarray}}
\newcommand{\eeq}{\end{eqnarray}}

\def\nn{\nonumber}
\def\fft#1#2{{#1 \over #2}}

\newcommand{\SO}{\mathrm{SO}}
\newcommand{\U}{\mathrm{U}}

\newcommand{\RR}{\mathbb{R}}

\def\ha{\hat{a}}
\def\hb{\hat{b}}
\def\tchi{\tilde{\chi}}

\def\CJ{{\cal J}}

\def\CS{{\cal S}}

\newcommand{\bbibitem}[1]{\bibitem{#1}\marginpar{#1}}

\newcommand{\ads}[1]{AdS$_{#1}$}
\newcommand{\cft}[1]{CFT$_{#1}$}
\newcommand{\sph}[1]{S$^{#1}$}

\def\Label#1{\label{#1}%
  \smash{\hbox to0pt{\raise1ex\hbox{\tiny[#1]}\hss}}}
\def\noLabels{\let\Label=\label}
\def\nobbibitem{\let\bbibitem=\bibitem}

\begin{document}

\begin{titlepage}

\begin{flushright}\vspace{-3cm}
{\small
%{\tt arXiv:yymm.nnnn} \\
 IPM/P-2012/048 \\
\today }\end{flushright}
%\vspace{-.5cm}

\begin{center}
{{\Large{\bf{Near-Extremal Vanishing Horizon AdS${}_5$ Black Holes and Their CFT Duals}}}} \vspace{5mm}

{\large{{\bf Maria Johnstone,\footnote{e-mail:
M.J.F.Johnstone@sms.ed.ac.uk}$^{,a}$ M.M. Sheikh-Jabbari\footnote{e-mail:
jabbari@theory.ipm.ac.ir}$^{,b}$,
Joan Sim\'on\footnote{e-mail:
j.simon@ed.ac.uk}$^{,a}$ and \\ H. Yavartanoo\footnote{e-mail:
yavar@khu.ac.kr}$^{,d}$  }}}
\\

\vspace{5mm}

\bigskip\medskip
\begin{center}
%\centerline
{$^a$ \it School of Mathematics and Maxwell Institute for Mathematical Sciences,\\
King's Buildings, Edinburgh EH9 3JZ, United Kingdom}\\
\smallskip
%\centerline
{$^b$ \it School of Physics, Institute for Research in Fundamental
Sciences (IPM),\\ P.O.Box 19395-5531, Tehran, Iran}\\
\smallskip
{$^c$ \it  Department of Physics, Kyung Hee University, Seoul 130-701, Korea}\\
\end{center}
\vfil

\end{center}
\setcounter{footnote}{0}

%%%%%%%%%%%
\begin{abstract}
\noindent
We consider families of charged rotating asymptotically AdS${}_5$ Extremal black holes with Vanishing Horizon (EVH black holes) whose near horizon geometries develop locally AdS${}_3$ throats. Using the AdS${}_3$/CFT${}_2$ duality, we propose an EVH/CFT$_2$ correspondence to describe the near-horizon low energy IR dynamics of near-EVH black holes involving a specific large $N$ limit of the 4d ${\cal N}=4$ SYM. We give a map between the UV and IR near-EVH excitations, showing that the `UV first law' of thermodynamics reduces to the `IR first law' satisfied by the near horizon BTZ black holes in this near-EVH limit. We also discuss the connection between our EVH/CFT proposal and the Kerr/CFT correspondence in the cases where the two overlap.

\end{abstract}

%\vspace{0.5in}

\end{titlepage}
%%%%%%%%%%%%% END OF TITLE PAGE %%%%%%%%%%%%%%%%%%%%%%%%%%

\renewcommand{\baselinestretch}{1.05}  %Line spacing

\tableofcontents

\newpage

%%%%%%%%%%%
\section{Introduction}
%%%%%%%%%%%

There has been a lot of progress in the microscopic understanding of black hole thermodynamics in some special cases within string theory involving extremal and near-extremal black holes/branes following the seminal work by Strominger and Vafa \cite{Strominger:1996sh}.

It was later appreciated that the appearance of an AdS${}_3$ throat in the near-horizon region of many of these black holes is at the heart of this microscopic description \cite{Strominger:1997eq}, using the seminal work of Brown and Henneaux \cite{Brown-Henneaux} establishing the existence of two Virasoro algebras as the asymptotic symmetry algebra of asymptotically AdS${}_3$ space-times. This was eventually properly understood in the context of the AdS/CFT correspondence \cite{Maldacena:1997re}.

More recently, holographic descriptions for generic extremal black holes with AdS${}_2$ throats in their near horizons were proposed. These involved either the AdS${}_2$/CFT${}_1$ correspondence \cite{Sen-AdS2/CFT1}  or 2d chiral CFTs following the so called Kerr/CFT correspondence \cite{Kerr-CFT}. Whenever the horizon of the black hole is compact, both proposals suggest the existence of a {\it non-dynamical} dual description, in the sense of not allowing finite energy excitations, whose vacuum degeneracy accounts for the macroscopic black hole entropy.

In this work, we continue our previous investigations of (near-)Extremal Vanishing Horizon (near-EVH) asymptotically AdS${}_5$ black holes \cite{Balasubramanian:2007bs,Fareghbal:2008ar,static case} by adding rotation in 5d, using the family of black holes constructed in \cite{Chong:2005da}. In a generic near-EVH black hole, by definition,   vanishing of the horizon area  appears because a one dimensional cycle on the horizon manifold becomes of zero size, so that its near-horizon geometry includes a 3d metric of the form\footnote{Many different examples of EVH black holes have been studied in the literature, e.g. see \cite{KKEVH,EVH-examples,Berkooz:2012qh} and explicitly checked that the appearance of a local \ads{3} is generic.}
\begin{equation}
  ds^2\propto -\epsilon^2\rho^2\frac{d\tau^2}{\ell_3^2} + {\ell_3^2}\frac{d\rho^2}{\rho^2} + \epsilon^2\rho^2d\varphi^2 + \dots\label{near-EVH-generic}
\end{equation}
The vanishing cycle is responsible for both the vanishing of the entropy and the transformation of the standard AdS${}_2$ throat into a local AdS${}_3$ throat, providing a bridge between the two successful theoretical scenarios mentioned above.

As explained in \cite{Massless-BTZ}, the geometry \eqref{near-EVH-generic} appears in the near core geometry of massless BTZ black holes. The latter has two inequivalent near horizon limits: one giving rise to the null self-dual orbifold, corresponding to the vacuum of the chiral sector surviving the extremal limit; and a second giving rise to the {\it pinching} \ads{3} orbifold, \ads{3}$/\mathrm{Z}_{1/\epsilon}$ with $\epsilon\to 0$, which decouples both chiral sectors by sending the boundary cylinder to zero. The null self-dual orbifold is the vacuum of a 2d chiral theory in which {\it finite} chiral excitations exist, giving rise to different DLCQ sectors of the full non-chiral CFT \cite{Balasubramanian:2009bg,Balasubramanian:2010ys}.  On the other hand, at least when the entropy vanishes, the pinching orbifold resembles the situation within the AdS${}_2$/CFT${}_1$ and Kerr/CFT descriptions.

As first mentioned in \cite{Massless-BTZ} and further elaborated in \cite{static case,KKEVH}, the conclusion about the absence of dynamics in a 2d CFT on a pinching orbifold cylinder, i.e. a cylinder of radius $R\to 0$, can be avoided if the central charge $c$ of the theory is sent to infinity keeping $cR=\text{finite}$. This can be understood in a very intuitive way \cite{Massless-BTZ}: the energy levels of a 2d CFT on a cylinder of radius $R$ are of the form ${\cal E}/R$, where ${\cal E}$ is a typical eigenvalue of $L_0, \bar L_0$, whereas the central charge $c$ controls the mass gap (the differences between the ${\cal E}$ eigenvalues are of the form $\hat{E}/c$).  Altogether the energy separation between the energy levels is given by $\hat{E}/(cR)$. So, the energy eigenvalues of a 2d CFT of central charge $c$ on a cylinder of radius $R$ is the same as those of a 2d CFT with central charge $\tilde{c}=cK$, on the same cylinder orbifolded by $Z_K$ (the latter may be viewed as a cylinder of radius $\tilde{R}=R/K$).

Recalling the Brown-Henneaux formula for the central charge $c=\frac{3\ell_3}{2G_3}$ \cite{Brown-Henneaux}, the $c\to\infty$ scaling in the \cft{2} can be realised in gravity by a $G_3\to 0$ limit, keeping the \ads{3} radius $\ell_3$ fixed. In the geometry \eqref{near-EVH-generic}, this would correspond to $G_3\sim\epsilon\to 0$. In higher dimensions, $d$ dimensional (near)-EVH \emph{black holes}, with {\it compact} $d-2$ dimensional horizons, would require $G_d\sim \epsilon \to 0$.

All these considerations motivated the {\it near-EVH triple decoupling} limit introduced in \cite{KKEVH}. This defines near-EVH black holes as those with vanishing horizon area $A_{\text{h}}$ and Hawking temperature $T_{\text{H}}$ such that
\be\label{triple-scaling}
A_{\text{h}}\sim T_{\text{H}}\sim G_d\to 0,\qquad A_{\text{h}}/G_d=\text{fixed},\ \ T_{\text{H}}/G_d=\text{fixed}\,.
\ee
In this limit the dynamics of the black hole is described by a 2d CFT with a finite central charge, temperature proportional to $T_{\text{H}}/G_d$ and entropy {\it equal to} $A_{\text{h}}/(4G_d)$. At the level of the near-horizon geometry this corresponds to a generic BTZ black hole.

A further feature of our setups, which is not generic to all EVH black holes,  is the existence of a UV description in terms of ${\cal N}=4$ SYM. The appearance of non-trivial IR 2d CFTs describing specific sectors of this theory has been explored in \cite{Balasubramanian:2007bs,Fareghbal:2008ar,static case} and provides non-trivial evidence regarding the rearrangement of field theory degrees of freedom in these sectors based on gravity considerations. The scaling of Newton's constant \eqref{triple-scaling} implies we will be focusing on a large $N$ limit of the ${\cal N}=4$ SYM in which $N^2T_{\text{H}}$ is kept fixed.

This paper is organized as follows. In section \ref{sec:pbh}, we introduce the charged rotating asymptotically AdS${}_5$ black holes we shall analyse, reviewing their charges and thermodynamics. In section \ref{sec:EVH}, we determine the set of EVH and near-EVH black holes within this family. In section \ref{sec:nh}, we study the
near-horizon geometries of these near-EVH black holes. In section \ref{sec:evhcft}, we compute the central charges of the IR 2d CFTs, map the UV ${\cal N}=4$ SYM quantum numbers to the IR 2d CFT excitations and derive the IR first law of thermodynamics from the corresponding one at the UV. In section \ref{sec:kerrcft}, we discuss the relation between the EVH/CFT and the Kerr/CFT correspondences within this class of black holes. The last section is devoted to concluding remarks.  Some technical details of computations have been gathered in the Appendices.

%%%%%%%%%%%%%%%%%%%%%%%%%%
\section{Charged rotating AdS${}_5$ black holes}
\label{sec:pbh}
%%%%%%%%%%%%%%%%%%%%%%%%%%

The particular class of black holes considered here are solutions to the $\U(1)^3$ 5d gauged supergravity whose bosonic action is  \cite{Gunaydin:1984ak}
\bea
\label{action5d}
S_{5d}=\frac{1}{16\pi G_5} \int d^5x \sqrt{-g}&&\hspace{-5mm}\bigg(R-\frac{1}{2}\partial\vec{\varphi}^2 -\frac{1}{4}\sum_{i=1}^3 X_i^{-2} F_i^2 +\frac{4}{l^2}\sum_{i=1}^3 X_i^{-1} \nonumber \\
&&+\frac{1}{4}\epsilon_{\mu\nu\rho\sigma\lambda}F_1^{\mu\nu}  F_2^{\rho\sigma} A^{\lambda}_3 \bigg),
\eea
where $\vec{\varphi}=(\varphi_1, \varphi_2)$ and
\be
X_1=e^{-\frac{1}{\sqrt{6}}\varphi_1-\frac{1}{\sqrt{2}}\varphi_2},\quad X_2=e^{-\frac{1}{\sqrt{6}}\varphi_1+\frac{1}{\sqrt{2}}\varphi_2},\quad X_3=e^{\frac{2}{\sqrt{6}}\varphi_1}\,.
\ee
Its most general asymptotically \ads{5} black hole solutions include three electric charges, two angular momenta (spins) and mass. They form a six parameter family of solutions. The four parameter subclass of static black holes was constructed in \cite{Behrndt:1998jd}. The EVH black holes in this subclass, which are two-charge black holes, were studied in \cite{Balasubramanian:2007bs,Fareghbal:2008ar, static case}.

In this work we consider black holes with two independent spins, mass and two equal R-charges, with the third R-charge a function of the remaining charges. These solutions were first constructed in \cite{Chong:2005da}\footnote{The most general six-parameter solution to this theory was constructed in \cite{most-general}.}. Their metrics are
\bea\label{5d BH metric}
ds^2&=&\,H^{-\frac{4}{3}}\,\Big [ -\frac{X}{\rho^2}\,
  (dt-a\,\sin^2\theta\,\frac{d\phi}{\Xi_{a}}-
   b\,\cos^2\theta\,\frac{d\psi}{\Xi_{b}})^2\nn\\
&&+\frac{C}{\rho^2}(\,\frac{ab}{f_3}dt-\frac{b}{f_2}
   \sin^2\theta \frac{d\phi}{\Xi_{a}}-
   \frac{a}{f_1}\cos^2\theta \frac{d\psi}{\Xi_{b}})^2 \\
%%%
&+&\frac{Z\sin^2\theta}{\rho^2}(\frac{a}{f_3}dt-
   \frac{1}{f_2}\frac{d\phi}{\Xi_{a}})^2+
  \frac{W\cos^2\theta}{\rho^2}(\frac{b}{f_3}dt-
   \frac{1}{f_1}\frac{d\psi}{\Xi_{b}})^2\Big ]+
H^{\frac{2}{3}}\,(\frac{\rho^2}{X}dr^2+\frac{\rho^2}{\Delta_\theta}d\theta^2\,)\,, \nn
\end{eqnarray}
gauge and scalar fields
\begin{eqnarray}
A^1&=&A^2= P_1(dt\,-\,a\,\sin^2\theta\,
  \frac{d\phi}{\Xi_{a}}\,-\,b\cos^2\,\theta\,\frac{d\psi}{\Xi_{b}})\nn\\
%%%
A^3&=& P_3(b\,\sin^2\theta\,\frac{d\phi}{\Xi_{a}}\,+\,a\,
  \cos^2\theta\,\frac{d\psi}{\Xi_{b}})\nn\\
%%%
X_1&=&X_2=H^{-\frac{1}{3}},\quad X_3=H^{\frac{2}{3}}\,,
\eeq
where
\bea\label{d5sol}
H&=&\tilde \rho^2/\rho^2,\quad \rho^2=r^2\,+\,a^2\,\cos^2\theta\,
   +\,b^2\,\sin^2\theta,\quad \tilde \rho^2=\rho^2\,+\,q\,,\nn\\
%%%%
f_1&=&a^2\,+\,r^2,\quad f_2=b^2\,+\,r^2,\quad f_3=(a^2+r^2)(b^2+r^2)\,+
   qr^2;\nn\\
%%%
\Delta_\theta &=&1-\frac{a^2}{\ell^2}\,\cos^2\theta-\frac{b^2}{\ell^2}\,\sin^2\theta,\nn\\
%%%
X(r)&=&\fft{1}{r^2}(a^2+r^2)(b^2+r^2)-2m+(a^2+r^2+q)(b^2+r^2+q)/\ell^2
 \,,\\
%%%
C&=&f_1\,f_2(X\,+\,2m\,-q^2/\rho^2),\nn\\
%%%
Z&=&-b^2\,C\,+\,\fft{f_2\,f_3}{r^2}[ f_3\,-\frac{r^2}{\ell^2}\,
             (a^2\,-b^2)(a^2+r^2+q)\cos^2\theta],\nn\\
%%%
W&=&-a^2\,C\,+\,\fft{f_1\,f_3}{r^2}[ f_3\,+
  \frac{r^2}{\ell^2}\,(a^2\,-b^2)(b^2+r^2+q)\sin^2\theta]\,,\nn\\
\Xi_{a} &=& 1 -\frac{a^2}{\ell^2}\,,\qquad \Xi_{b}= 1 - \frac{b^2}{\ell^2},  ,\qquad P_1=\fft{\sqrt{q^2+2mq}}{\tilde \rho^2}\,,\qquad	P_3=\fft{q}{\rho^2}     \nn
\eea
This family of solutions is specified by four parameters $(a,b,q,m)$. The metric \eqref{5d BH metric} is written in an asymptotically {\it rotating} frame (ARF). The coordinate transformation
\begin{equation}
  \phi^S = \phi + \frac{a}{\ell^2}\,t\,, \quad \quad \psi^S = \psi + \frac{b}{\ell^2}\,t
\end{equation}
brings it to the {asymptotically} {\it static} frame (ASF), where it is manifestly the \ads{5} metric with radius $\ell$ in the standard global coordinate system.

\paragraph{Charges and thermodynamics:} Whenever the equation $X(r)=0$ allows real solutions, the configurations \eqref{5d BH metric} describe a family of black holes. This is discussed in detail in appendix \ref{ap:horizon} in the regime of charges we are interested in this work. In the following, we review their charges and thermodynamics.

The angular momenta and electric charges of the family of black hole solutions can be evaluated using Komar and Gaussian integrals respectively \cite{Chow:2008dp},
\bea
J_{a} &=& \frac{\pi  a\,  (2m + q\, \Xi_{b})}{4G_5 \Xi_{b}\, \Xi_{a}^2}\,,\; \qquad
J_{b} = \frac{\pi   b\,  (2m +  q \, \Xi_{a})}{4G_5 \Xi_{a}\, \Xi_{b}^2}\,,\label{eq:angmom}\\
Q_1&=& Q_2= \frac{\pi  \sqrt{q^2+2mq}}{4G_5\Xi_{a}\, \Xi_{b}}\,,\quad
Q_3 = -\frac{\pi a b q  }{4G_5 \ell^2\Xi_{a}\, \Xi_{b}}\,.\label{eq:rcha}
\eea
Note that $J_a=J_\phi$ and $J_b=J_\psi$ are the standard angular momentum associated with rotations along the $\phi$ and $\psi$ angles in the 3-sphere. The horizon structure determines the thermodynamic properties of the black hole. Its temperature can be computed through the horizon surface gravity, leading to
\bea
\label{eq:th}
T_{\textrm{H}} =
\frac{2r_+^6
 +  r_+^4 (\ell^2 + a^2 + b^2 + 2 q)- a^2 b^2\ell^2}{2 \pi r_+ \ell^2[(r_+^2 +
a^2) (r_+^2 + b^2) + q r_+^2]}\,,
\eea
whereas the Bekenstein--Hawking entropy is proportional to the area of the black hole horizon,
 \begin{equation}
S_{\text{BH}} = \frac{\pi^2 [(r_+^2 + a^2)
(r_+^2 + b^2) + q r_+^2]}{2 G_5 \Xi_a \Xi_b r_+}\,.\label{eq:ts}
\end{equation}

The outer horizon $(r=r_+)$ is the Killing horizon generated by the Killing vector field $K=\partial_t +\Omega^S_a \partial_{\phi}+\Omega^S_b\partial_{\psi}$, where $\Omega^S_a$ and $\Omega^S_b$ stand for the angular velocities on the horizon in the ASF \cite{Chong:2005da}
\begin{equation}\label{ang velocities in ASF}
\Omega^S_a=\frac{a(r_+^4+r_+^2b^2+r_+^2q+\ell^2b^2+\ell^2r_+^2)}{\ell^2(a^2+r_+^2)(b^2+r_+^2)\,+  \ell^2 qr_+^2}, \,\,\,\,\,
\Omega^S_b=\frac{b(r_+^4+r_+^2a^2+r_+^2q+\ell^2a^2+\ell^2r_+^2)}{\ell^2(a^2+r_+^2)(b^2+r_+^2)\,+  \ell^2 qr_+^2}.
\end{equation}
The same angular velocities in the ARF equal $\Omega^R_a = \Omega^S_a-\frac{a}{\ell^2}$ and $\Omega^R_b = \Omega^S_b-\frac{b}{\ell^2}$.

The electrostatic potentials $\Phi_I$ associated to the electric charges can be computed through the definition
$\Phi_I=-K^{\mu}A^I_{\mu}|_{r=r_+}$. One finds
\bea
 \Phi_1=\Phi_2 = \frac{\sqrt{q^2+2mq}\;r_+^2 }{(a^2+r_+^2)(b^2+r_+^2)\,+   qr_+^2},\qquad
\Phi_3 = \frac{qab}{(a^2+r_+^2)(b^2+r_+^2)\,+   qr_+^2} \label{eq:5dchepot}\,.
\eea

The mass of these black holes can be determined either from the ADM mass or by integrating the first law of thermodynamics
 \begin{equation}
T_{\text{H}}\,dS_{\text{BH}} = dE - \Omega_a\,dJ_a - \Omega_b\,dJ_b - \sum_{i=1}^3 \Phi_i\,dQ_i,
 \label{eq:UVflaw}
\end{equation}
giving rise to
\begin{equation}
E= \frac{\pi [2m(2\Xi_{a} + 2\Xi_{b} - \Xi_{a}\, \Xi_{b}) +
                 q (2\Xi_{a}^2 + 2\Xi_{b}^2 + 2\Xi_{a}\, \Xi_{b}
      - \Xi_{a}^2\, \Xi_{b} - \Xi_{b}^2\, \Xi_{a})\ ]}{8 G_5 \Xi_{a}^2\, \Xi_{b}^2}\,.\label{eq:mass}
\end{equation}

There are BPS black holes among the family \eqref{5d BH metric} with energy $E$ a linear combination of the electric charges and angular momenta \cite{Chong:2005da,More-on-AdS5-bh,Chow:2008dp}. The generic BPS solutions in this class preserve 1/4 of the supersymmetries of the original theory, i.e. two (out of eight) real supercharges \cite{Chow:2008dp,korean}. We also note that BPS \ads{5} black holes can only happen if we have electric charges turned on, i.e. neutral rotating \ads{5} black holes cannot be BPS; moreover, all the static BPS \ads{5} black holes have naked singularities, which can be removed by the addition of rotation \cite{korean}.

\paragraph{Embedding in type IIB supergravity:}  All solutions to \eqref{action5d} can be uplifted to on-shell 10 dimensional type IIB supergravity solutions using the ansatz \cite{Cvetic:1999xp}
\begin{equation}\label{eq:d10embed}
 ds_{10}^2=\sqrt{\widetilde{\Delta}}\,ds_5^2+\frac{\ell^2}{\sqrt{\widetilde{\Delta}}}\displaystyle\sum_{i=1}^{3}X_i^{-1}(d\mu_1^2+\mu_i^2(d\psi_i+A^i/\ell)^2),
\end{equation}
with $X_i$ as in \eqref{d5sol}, $\mu_i$ are functions parameterizing a unit 2-sphere
\begin{equation}\label{mu-i}
\mu_3=\cos{\alpha} \,,\qquad \mu_2=\sin{\alpha}\sin{\beta}\,,\qquad \mu_1=\sin{\alpha}\cos{\beta}\,, \,\,\,\, \alpha,\beta \in [0,\frac{\pi}{2}]
\end{equation}
and
\begin{equation}\label{Delta-tilde}
\widetilde{\Delta} = \displaystyle\sum_{i=1}^{3}X_i\mu_i^2=H^{-\frac{1}{3}}(\mu_1^2+\mu_2^2)+H^{\frac{2}{3}}\mu_3^2\,.
\end{equation}
The 10d metric \eqref{eq:d10embed} is a solution of IIB supergravity with constant dilaton $\phi$, where $e^\phi=g_s$, and selfdual RR-fiveform field \cite{Cvetic:1999xp}.

Recalling that $G_{10}=8\pi^6g_s^2\ell_s^8$ and performing a standard compactification over the 5-sphere, we learn that the 5d Newton's constant equals
\be\label{G5-G10}
G_5=G_{10} \frac{1}{\pi^3 \ell^5}=\frac{\pi}{2} \frac{\ell^3}{N^{2}}\,,
\ee
where $N$ is the 5-sphere RR five-form flux and $\ell$ its radius $\left(\ell^4=4\pi g_s\ell_s^4\ N\right)$. Thus, all black hole charges scale like $N^2$.

The 10d perspective allows to reinterpret the 5d electrostatic potentials $\Phi_I$ and electric charges $Q_I$ as angular velocities $\Omega_i$ and angular momenta $J_i$ on the transverse S${}^5$. Due to the conventions in
\eqref{eq:d10embed}, their relation is
\be\label{Omega-S5-Phi}
\Omega_i=-\frac{\Phi_i}{\ell}\,,\qquad J_i=-\ell Q_i\,.
\ee
As in 5d, all 10d angular velocities can be computed by requiring the vanishing of the norm of the Killing vector field $K=\partial_t +\Omega_a^S\partial_{\phi^S} + \Omega_b^S\partial_{\psi^S} +\Omega_i\partial_{\phi_i}$ at the outer horizon.

\paragraph{Four dimensional ${\cal N}=4$  SYM description:} Using AdS/CFT, the black holes \eqref{5d BH metric} correspond to thermal states in the dual ${\cal N}=4$ SYM defined on $\RR\times$\sph{3} carrying charges :
\be\label{SYM-charges}
\begin{split}
\Delta&=\ell{E}\,,\qquad  \CJ_1=\CJ_2=Q_1=\frac{\sqrt{q^2+2mq}}{2\ell^2\Xi_a\Xi_b} N^2\,,\\
\CS_a =J_a&=\frac{a(2m+q\Xi_b)}{2 \ell^3\Xi_a^2\Xi_b}N^2\,,\qquad \CS_b=J_b=\frac{b(2m+q\Xi_a)}{2 \ell^3\Xi_b^2\Xi_a}N^2\,.
\end{split}
\ee
As usual, energy $E$ becomes conformal dimension $\Delta$, angular momenta $J_a,\,J_b$ $\SO(4)$ spins ${\cal S}_a,\ {\cal S}_b$ and the electric charges $Q_i$, R-charges $\CJ_i$. By construction, these are functions of the the four parameters $(a,b,q,m)$ and scale like $N^2$.

\paragraph{Intersecting giant description:}

 As a 10 dimensional solution the metric corresponds to a system of two stacks of intersecting black giant branes. These are topologically three-sphere branes with their world-volume on two three cycles on the \sph{5}, intersecting on a circle parameterized by $\psi_3$ in our coordinates. These giant gravitons are rotating on $\psi_1$ and $\psi_2$ directions which are transverse to their world-volume; the electric charges correspond to the angular momenta of the giants in each stack.
The number of giants in each stack is then given by \cite{Myers:2001aq}
\be
N_1=N_2=\frac{2\CJ_1}{N}\,.
\ee
As we see for generic values of the $a,b,q$ parameters the number of giants in each stack grows as $N$, and hence we are dealing with a system of order $N^2$ giants.

We should comment that the intersecting giant graviton picture is a good one if we are close to a BPS point, where moving slightly away from that would correspond to turning on excitations (deformations) on the giant graviton world-volume. In this picture the spins $J_a,\ J_b$ would correspond to rotating  the giant on the \sph{3}$\in$\ads{5}. (Recall that each giant is already rotating on a circle in \sph{5}.)
%{\bf I am still not sure whether we should keep this; references ? check Maldacena; quote Suvrat's unpublished thesis %?}

%%%%%%%%%%%%%%%%%%%
\section{The set of EVH black holes}
\label{sec:EVH}
%%%%%%%%%%%%%%%%%%%

Consider the four dimensional black hole parameter space, either in terms of $(a,b,q,m)$ or $(a,b,q,r_+)$.
We are physically defining the subset of EVH configurations as a limit of near-extremal black holes in which $S_{\text{BH}}\,,T_{\text{H}}\sim \epsilon\to 0$, keeping $S_{\text{BH}}/T_{\text{H}}$ {\it finite}. Inspection of \eqref{eq:th} and \eqref{eq:ts} reveals
\be\label{EVH-def}
S_{\text{BH}},T_{\text{H}}\sim \epsilon \to 0 \quad\Rightarrow \quad r_+\sim \epsilon\,,\quad {a}\sim \epsilon^\alpha\,,\quad {b}\sim \epsilon^{\beta},\ \beta\geq \alpha\geq 0,\ \alpha+\beta\geq 2\,.
\end{equation}
where we used the $a\leftrightarrow b$ exchange symmetry to assume $b\leq a$. Thus, EVH black holes require
\begin{equation}
\boxed{r_+=0 \quad  \text{and} \quad  ab=0\,.}
\end{equation}
These describe a \emph{bifurcate hypersurface}, corresponding to the $a=0$ or $b=0$ branches. Since $b\leq a$, we shall focus on the $b=0$ branch. Requiring $X(r_+=b=0)$ to vanish, ensuring the presence of a horizon, gives rise to the further constraint
\be\label{HX2}
m=\frac{q^2+{a}^2(\ell^2+q)}{2\ell^2}\,.
\ee
In the following, we shall distinguish between two types of EVH configurations:
\be
\label{EVHgen}
{\rm{\bf Static:}} \quad {a}={b}=r_+=0\quad\quad \mathrm{and}\quad\quad  {\rm{\bf Rotating:}} \quad {b}=r_+=0,\;\; {a}\neq 0\,.
\ee
\begin{figure}[htb]
 \begin{center}
\includegraphics[scale=0.55]{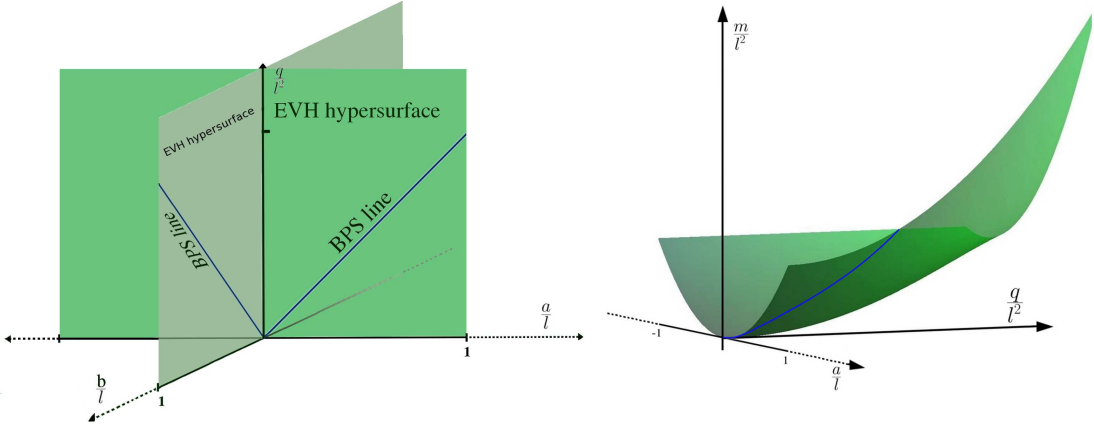}
\caption{\emph{Bifurcate EVH hyperplane}. The left figure shows the EVH hyperplane in $(a,b,q)$ space. The $a=b=0$ $q$ axis stands for static EVH  black holes, while $b=0$, $a\neq 0$ points correspond to rotating EVH black holes. The subset of EVH BPS configurations, which occur for $q=\ell a,\ b=0$ or $q=b\ell, a=0$, are indicated by $45^\circ$ lines. The right figure shows the $b=0$ branch of the EVH hyperplane in $(a,q;m)$ space. }
\label{EVH}
\end{center}
\end{figure}
In figure \ref{EVH}, we illustrate the space of EVH configurations, where we already took into account that $q\geq 0$ and $|a|,|b|\leq \ell$. Static configurations correspond to the $q$-axis at $a=0$ with $2m=q^2/\ell^2$, whereas rotating ones correspond to generic $a\neq 0$ points with $m$ given by \eqref{HX2}. Among the rotating BPS EVH black holes we have quarter-BPS configurations specified by $q=a \ell,\ b=0$, or $q=b \ell,\ a=0$.

%%%%%%%%%%%%%%%%%%%
\subsection{Near-EVH black holes}
\label{sec:near-evh}
%%%%%%%%%%%%%%%%%%%

To explore the physics of near-EVH black holes, we describe regions in parameter space close to the EVH hypersurface.
Given a generic $b=0$ EVH point parameterized by \eqref{HX2}, one can decompose the space of deformations into
{\it tangential} and {\it orthogonal}. The first correspond to
\begin{equation}
m+\delta m=\frac{(q+\delta q)^2+{(a+\delta a)}^2(\ell^2+q+\delta q)}{2\ell^2}\,.
\label{eq:tdef}
\end{equation}
These tangential deformations take us  from one EVH black hole to a different one on the EVH hyperplane. Orthogonal deformations correspond to excitations of an EVH black hole, giving rise to near-EVH black holes. We will study the static and rotating cases separately.

\paragraph{Near-EVH static black holes:} Static EVH  configurations are described by
$(a,b,q;m)=(0,0,q;q^2/2\ell^2)$ and hence the most general orthogonal deformation is
\begin{equation}
\label{eq:nevhstatic}
\left(0,0,q;\frac{q^2}{2\ell^2}\right)\qquad \longrightarrow \qquad \left(\delta a, \delta b, q-\frac{q}{\ell^2}\delta m ;\frac{q^2}{2\ell^2}+\delta m\right)\,.
\end{equation}
Recalling \eqref{EVH-def}, one can choose $\delta m\sim \epsilon^2,\ \delta a\sim \epsilon^{\alpha}\,,\delta b\sim \epsilon^{\beta}$ where $\beta\geq\alpha> 0,\ \alpha+\beta\geq 2$. Demanding to have a black hole, i.e. $X(r)=0$ to have real positive solutions, implies $\alpha\geq 1$. One may hence choose,
\be\label{near-EVH-static}
\delta m=M\epsilon^2\,,\quad \delta a=\ha \epsilon\,,\quad \delta b=\hb\epsilon\,,
\ee
where $\ha,\ \hb$ are either finite or may go to zero in some positive power of $\epsilon$. Thus, the most general physical excitations of static EVH black holes are described by the three parameters $M,\ha,\hb$. This is consistent with the co-dimension three of the the EVH surface describing the static case \eqref{EVHgen}, as depicted in figure \ref{EVH}.

Once these deformations are turned on, the equation determining the horizon location $X(r_\pm)=0$ becomes
\be\label{X(r)-Static-like}
\mathbf{V}_s\left(\frac{r}{\epsilon}\right)^4- \left[2\mathbf{W}_sM-\mathbf{Y}_s(\ha^2+\hb^2)\right]\left(\frac{r}{\epsilon}\right)^2+\ha^2\hb^2=0\,,
\ee
with
\be\label{Vs-Ws-Ys}
\mathbf{V}_s=1+\frac{2q}{\ell^2}\,,\qquad \mathbf{W}_s=1+\frac{q^2}{\ell^4}\,,\qquad \mathbf{Y}_s=1+\frac{q}{\ell^2}.
\ee
This ensures both $r_\pm\sim \epsilon$ as required in \eqref{EVH-def}. In this limit one can work out the temperature \eqref{eq:th} and entropy \eqref{eq:ts}
\be\label{entropy-temp-static-like}
S_{\text{BH}}=\pi N^2 \frac{q}{\ell^2} \frac{r_+}{\ell}\,,\qquad T_{\text{H}}=\frac{(1+\frac{2q}{\ell^2})-\frac{\ha^2\hb^2}{r_+^4}}{2\pi q} r_+=\frac{\mathbf{V}_s}{2\pi q}\frac{r^2_+-r^2_-}{r_+}\,,
\ee
where we used $r_+^2r_-^2=\mathbf{V}_s^{-1} a^2b^2$, which follows from \eqref{X(r)-Static-like}. By construction,
 $S_{\text{BH}}\sim T_{\text{H}}\sim r_\pm\sim \epsilon$.

\paragraph{Near-EVH rotating black holes:} Using $m(a,q)$ in \eqref{HX2}, rotating EVH  configurations are described by $(a,0,q;m(a,q))$. Their most general orthogonal deformation is hence
\be\label{near-EVH-KK-like}
(a, 0, q; m )\quad \longrightarrow \quad \left(a-a\left(1+\frac{q}{\ell^2}\right)\frac{\delta m}{\ell^2}, \delta b, q-\left(q+\frac{1}{2}a^2\right)\frac{\delta m}{\ell^2}; m+\delta m \right)\,.
\ee
For finite $a,q$, the scaling \eqref{EVH-def} requires
\be\label{delta b-delta M}
\delta b= \hb\epsilon^2\ , \quad \delta m= M\epsilon^2\,.
\ee
Unlike the static EVH case, the most general excitations of a rotating EVH black hole are described by two parameters, $M, \hb$. This is consistent with the co-dimension two surfaces defining them in \eqref{EVHgen}, as shown in figure \ref{EVH}.

These deformation parameters determine the location of the horizons
\be\label{X(r)-KK-like}
\mathbf{V}\left(\frac{r}{\epsilon}\right)^4-2 \mathbf{W}M \left(\frac{r}{\epsilon}\right)^2+a^2\hb^2=0
\ee
with
\be\label{V0-W}
\mathbf{V}=1+\frac{2q}{\ell^2}+\frac{a^2}{\ell^2}, \qquad \mathbf{W}=1+\left(\frac{q}{\ell^2}+\frac{a^2}{2\ell^2}\right)^2+\frac{a^2}{\ell^2}\left(1+\frac{q}{\ell^2}\right)^2\,.
\ee
Notice its $a\to 0$ limit does {\it not} reproduce the static equation \eqref{X(r)-Static-like}. This is because the near-EVH scaling of $b$ with $\epsilon$ is different from the static case in \eqref{near-EVH-static} and, more importantly, because there exists a non-commuting order of limits between taking a near horizon limit and considering the $a\to 0$ limit, as we shall explicitly see in section \ref{sec:rot-nearevh}. For these reasons, we will study the static and rotating cases separately in the following, noting when the former can be obtained as a limit of the latter.

It is reassuring that whenever the parameters $\{a,q;\hb ,M\}$ allow real roots $r_\pm$\footnote{For a thorough analysis on when this occurs, see appendix \ref{ap:horizon}.}, the entropy and temperature of these near-EVH black holes equal
\be\label{S-T-NH-limit}
S_{\text{BH}}=\pi N^2 \frac{q+a^2}{\ell^2 \Xi_a} \frac{r_+}{\ell}\,,\qquad T_{\text{H}}=\frac{(1+\frac{2q+a^2}{\ell^2})-\frac{a^2\hb^2}{r_+^4}}{2\pi (q+a^2)} r_+=\frac{\mathbf{V}}{2\pi (q+a^2)} \frac{r^2_+-r^2_-}{r_+}\,.
\ee
By construction, the ratio $S_{\text{BH}}/T_{\text{H}}$ is finite. Notice these equations reproduce \eqref{entropy-temp-static-like} when written as a function of $r_\pm$ at $a=0$.

We stress expressions \eqref{S-T-NH-limit} resemble the analogue quantities for BTZ black holes, i.e.
$S_{\text{BTZ}}=\frac{\pi r_+}{2G_3}$ and $T_{\text{BTZ}}=\frac{r_+^2-r_-^2}{2\pi \ell_3^2 r_+}$. In fact, we will see in the coming sections that the near horizon of these configurations develops a locally AdS${}_3$ throat. The corresponding 3d Newton's constant $G_3$ and AdS${}_3$ radius $\ell_3$ will be such that this analogy will become an equality.

%%%%%%%%%%%%%%%%%%%%%%
\section{Near Horizon Geometry Analysis }
\label{sec:nh}
%%%%%%%%%%%%%%%%%%%%%%

In this section we study the near horizon geometries corresponding to the static and rotating EVH black holes identified in \eqref{EVHgen} together with their near-extremal versions \eqref{eq:nevhstatic}-\eqref{near-EVH-KK-like}.

%%%%%%%%%%%%%%%%
\subsection{Static EVH case}
%%%%%%%%%%%%%%%%

Let us consider a static EVH  black hole $(a=b=0)$ and study its deep interior geometry by expanding in small $\epsilon$ for $r=\epsilon\rho$. The metric expansion is
\begin{align}
\label{eq:a1}
ds^2&=\frac{q\ell^2 \mu_3}{\ell^2+2q}\bigg[-
\frac{ \mathbf{V}_s^2\epsilon^2}{q^2}  \rho^2 dt^2 +\frac{d\rho^2}{\rho^2}+ \frac{\ell^2 \mathbf{V}_s^2 \epsilon^2}{q^2}\rho^2d \psi_3^2   \bigg]  \\
\nonumber &+
q\mu_3 \bigg( d\theta^2 +\sin^2\theta d\phi^2
 + \cos^2\theta d\psi^2  \bigg)+
 \frac{\ell^2(d\mu_1^2+d\mu_2^2)}{\mu_3}
 \\ \nonumber
&+ \frac{\ell^2\mu_1^2}{\mu_3}\left( d\psi_1-\Omega^0_1 dt\right)^2   +\frac{\ell^2\mu_2^2}{\mu_3}\left( d\psi_2-\Omega^0_2 dt\right)^2\,,
\end{align}
Extremality determines the scaling $-\epsilon^2\rho^2\ dt^2$ together with $d\rho^2/\rho^2$ giving rise to an AdS${}_2$ throat responsible for the $\SO(2,1)$ isometry enhancement of the near horizon geometry  \cite{AAAA}.
The new feature here is the {\it vanishing} size of the one-cycle along $\psi_3$ as $\epsilon^2\rho^2$. Notice this is the isometry direction in the 5-sphere with vanishing R-charge $(J_3=0)$. This is responsible for the vanishing of the entropy and transforms the standard AdS${}_2$ throat into a local AdS${}_3$ throat, as discussed in the introduction. Even though $J_a=J_b=0$, the cycles along $\partial_\phi$ and $\partial_\psi$ remain of {\it finite} size.

The near horizon geometry is obtained by considering the limit
\be\label{coord-scaling-static-EVH}
\psi_i=\tilde{\psi}_i+\Omega^0_i\ t,\,\,(i=1,2) \quad t=\frac{\ell}{\sqrt{q}}\frac{\tau}{\epsilon}, \quad \psi_3=-\frac{\tilde{\chi}}{\epsilon}, \quad r = \epsilon\frac{\sqrt{q}}{\ell} x\,,
\ee
on the original black hole metric \eqref{eq:d10embed}, with
\be\label{omega-0-1}
\Omega^0_1=\Omega^0_2=-\frac{1}{\ell}\sqrt{1+\frac{q}{\ell^2}}=-\frac{\sqrt{\textbf{Y}_s}}{\ell}\,,
\ee
being the horizon angular velocity at the EVH point. The resulting metric is
\begin{align}
\label{NHa0b0}
 ds^2&=\mu_3\bigg[-
\frac{x^2d\tau^2}{\ell_3^2 }+\frac{\ell_3^2  dx^2}{x^2}+ x^2d \tilde{\chi}^2   \bigg] +   q\mu_3 \bigg( d\theta^2 +\sin^2\theta d\phi^2
 + \cos^2\theta d\psi^2  \bigg)
\nonumber \\
&+
 \frac{\ell^2}{\mu_3} (d\mu_1^2+d\mu_2^2+\mu_1^2 d\tilde{\psi}_1^2 +\mu_2^2 d\tilde{\psi}_2^2)\,.
\end{align}
Due to the $2\pi\epsilon$ periodicity in $\tilde{\chi}$, this geometry describes a warped locally AdS${}_3\times$S${}^3$ geometry, with radii given by
\be\label{AdS3-S3-radii}
R^2_{AdS_{3}}=\ell_3^2 = \frac{q}{\mathbf{V}_s} \,, \qquad R^2_{S^{3}}= q\,.
\ee
More precisely, the local AdS${}_3$ throat is the {\it pinching} AdS${}_3$ orbifold introduced in \cite{Massless-BTZ}, corresponding to the near horizon of a massless BTZ black hole\footnote{Massless BTZ black holes allow for a second near horizon limit giving rise to the {\it null self-dual} orbifold \cite{Massless-BTZ}. This structure also exists here. Defining $\tau_{IR}=(\sqrt{q}/\ell)\,t$, we can write the $\epsilon^2$ terms in \eqref{eq:a1} as $-\epsilon^2\left(d\tau_{IR}^2\ell_3^2-d\psi_3^2\right)$. Introducing light-like coordinates $x^\pm = \frac{\tau_{IR}}{\ell_3}\pm \psi_3$, the second near horizon limit implemented by $x^+=z^+$ and $x^-=\frac{z^-}{\epsilon^2}$ gives rise to the null self-dual orbifold. The two limiting geometries are related by an {\it infinite} boost.}. Once more, notice how the circle in AdS${}_3$ comes from the direction in the 5-sphere where there is no R-charge at the EVH point. Besides the pinching, which does not introduce a curvature singularity, the geometry \eqref{coord-scaling-static-EVH} is otherwise everywhere smooth except at $\mu_3=0$.

The static EVH case was already studied in \cite{Balasubramanian:2007bs,Fareghbal:2008ar,static case}. In these references, the near BPS EVH case $q\sim \epsilon \to 0$, which requires $N\to \infty$, was considered and argued to require a different limit in which the pinching is absent and the 4d ball decompactifies. Our analysis here shows the latter is not necessary, though the tree-level supergravity description becomes unreliable since both AdS${}_3$ and S${}^3$ become Planck scale.

%%%%%%%%%%%%%%%%%%%%
\subsection{Static near-EVH  case}
\label{sec:nevhstatic}
%%%%%%%%%%%%%%%%%%%%

Near-EVH static black holes are described in parameter space by \eqref{eq:nevhstatic}. These are excitations of the static EVH vacua. Hence, one expects them to be encoded in the near-horizon geometry as {\it pinching} BTZ black holes \cite{Massless-BTZ}. Indeed, as discussed in more detail in section \ref{IR-UV-section.5.2.2}, these excitations are described by mass and angular momentum of the pinching BTZ with the possible addition of constant electric and magnetic fields on the transverse 3-sphere, describing the $J_a$ and $J_b$ rotations.

These expectations are verified when we combine the near horizon limit \eqref{coord-scaling-static-EVH} with the angular coordinate redefinitions
\begin{equation}
\label{phi-psi-NEVH-static-scaling}
\tilde{\psi}={\psi}-\frac{\hb }{q}\frac{\ell}{\sqrt{q}}\tau\  - \frac{\ha \ell}{{q}} \tilde{\chi}\,,\quad  \tilde{\phi}={\phi}-\frac{\ha}{q}\frac{\ell}{\sqrt{q}} \tau - \frac{\hb \ell}{{q}}  \tilde{\chi}\,.
\end{equation}
The resulting near horizon metric is
\begin{align}
ds^2 &=\mu_3\bigg[ -\frac{(x^2- x_{+}^2)(x^2-x_{-}^2)}{ \ell_3^2 x^2}d\tau^2+\frac{\ell_3^2 x^2dx^2}{(x^2-x_{+}^2)(x^2-x_{-}^2)} + x^2(d\tilde{\psi_3}-\frac{x_+ x_-}{\ell_3x^2}d\tau)^2\bigg] \nonumber \\
 & +q\mu_3(d\theta^2+\sin^2\theta d\tilde{\phi}^2+\cos^2\theta d\tilde{\psi}^2 )+\frac{l^2}{\mu_3}(d\mu_1^2+d\mu_2^2+\mu_1^2d\tilde{\psi}_1^2+\mu_2^2d\tilde{\psi}_2^2)
\label{eq:hossein}
\end{align}
where $x_\pm$ are given in terms of \eqref{Vs-Ws-Ys} as
\begin{equation}\label{x-pm-btz-static}
x_\pm^2 =\frac{\ell^2}{2q\mathbf{V}_s}\left(2\mathbf{W}_sM-\mathbf{Y}_s(\ha^2+\hb^2)\pm\sqrt{\left(2\mathbf{W}_sM-\mathbf{Y}_s(\ha^2+\hb^2)\right)^2-4\mathbf{V}_s\hat{a}^2\hat{b}^2)}\right).
\end{equation}
Notice how the original pinching AdS${}_3$ turned into a pinching BTZ metric.\footnote{For the subset of excitations that preserve extremality, i.e. $x_+=x_-$, there exists a second near horizon limit again giving rise to the null self-dual orbifold.} The constant shifts in the AdS${}_3$ boundary coordinates $\{\tau,\, \tilde{\chi}\}$ in \eqref{phi-psi-NEVH-static-scaling} corroborate the existence of constant electric and magnetic fields responsible for the UV spins $J_a$ and $J_b$.

\paragraph{Temperature \& entropy:} To use the standard thermodynamic relations satisfied by BTZ black holes, we must compactify \eqref{eq:hossein} to three dimensions. Consider the ansatz
\begin{equation*}
ds^2=\mu_3\  g^{(3)}_{\mu\nu} dx^{\mu}dx^{\nu}
+q\mu_3(d\theta^2+\sin^2\theta d\tilde{\phi}^2+\cos^2\theta d\tilde{\psi}^2 )+\frac{l^2}{\mu_3}(d\mu_1^2+d\mu_2^2+\mu_1^2d\tilde{\psi}_1^2+\mu_2^2d\tilde{\psi}_2^2),
\end{equation*}
and plug it into the 10d type IIB supergravity action. Focusing on its Einstein-Hilbert term
\begin{equation*}
\frac{1}{16\pi G_{10}}\int  \sqrt{-g_{(10)}} \left( {}^{10}{\mathcal R} +\cdots\right) = \frac{1}{16\pi G_{3}}\int  \sqrt{-g_{(3)}} \left({}^{3} {\mathcal R} +\cdots\right)\,,
\end{equation*}
we can identify the 3d Newton's constant to be
\begin{equation}\label{3d-GN-static}
 \frac{1}{G_3}=\frac{q^{3/2}\ell^4}{16G_{10}}(2\pi)^4=\frac{2q^{\frac{3}{2}}N^2}{\ell^4}\,.
\end{equation}
Thus, the temperature and entropy of the pinching BTZ black holes \eqref{eq:hossein} equal
\begin{equation}\label{BTZ-temperature-entropy-static}
\begin{split}
T_{\text{BTZ}}&\equiv\frac{x_+^2-x_-^2}{2\pi x_+\ell_3^2}=\frac{\ell}{\epsilon\sqrt{q}}\ T_{\text{H}}\,,\\
S_{\text{BTZ}}&\equiv \frac{2\pi \epsilon\cdot x_+}{4G_3}=S_{\text{BH}}\,,
\end{split}
\end{equation}
where for $T_{\text{H}}$ and $S_{\text{BH}}$ we have used \eqref{entropy-temp-static-like}.
As expected, the second matches the original 10d black hole entropy whereas the first agrees with the scaling of time in \eqref{coord-scaling-static-EVH}. This confirms the expectations raised in section \ref{sec:EVH}, when interpreting the near-EVH temperature and entropy \eqref{entropy-temp-static-like} as BTZ thermodynamical quantities.

%%%%%%%%%%%%%%%%%%
\subsection{Rotating EVH case}
\label{section:NH-KK-EVH}
%%%%%%%%%%%%%%%%%%

Consider a rotating EVH black hole $(b=0)$ and study its deep interior geometry by expanding in small $\epsilon$ for $r=\epsilon\rho$. The metric expansion is
\begin{align}\label{NH-metric-no-chi-scale}
ds^2 &= \frac{h_1h_2}{a^2+q}\left[ -\frac{\mathbf{V}}{a^2+q}\epsilon^2\rho^2 dt^2 +\frac{a^2+q}{\mathbf{V}}
\frac{d\rho^2}{\rho^2} +\frac{a^2+q}{a^2}\rho^2\epsilon^2d\psi^2 \right] \nonumber \\
&+ \frac{(a^2+q) h_1h_2}{1-\frac{a^2}{\ell^2}\cos^2\theta}d\theta^2+\frac{\ell^2h_1}{h_2}(d\mu_1^2+d\mu_2^2) +\frac{a^2\ell^2\cos^2\theta}{(a^2+q) h_1h_2}d\mu_3^2
 +\frac{a^2\ell^2\cos^2\theta \mu_3^2}{h_1h_2}\left(d\psi_3+\frac{q}{a\ell}d\psi\right)^2 \nonumber \\
&+ \frac{\sin^2\theta\left[ a^2\mathbf{Y}_s-\frac{a^2}{\ell^2}(a^2+q)\cos^2\theta+q\Xi_a\mu_3^2   \right]}{\Xi_a^2h_1h_2}\left(d\phi -\Omega_a^{0R} dt     \right)^2      \nonumber \\
&-\frac{2a\ell\sqrt{ q\mathbf{Y}_s}\; \sin^2\theta}{\Xi_a h_1h_2 \sqrt{a^2+q}  }\left(d\phi -\Omega_a^{0R} dt     \right)\big[\mu_1^2\left(d\psi_1-\Omega^0_1 dt \right)+
\mu_2^2\left(d\psi_2-\Omega^0_2 dt \right)\big] \nonumber \\
&+\ell^2\frac{h_1\mu_1^2}{h_2} \left(d\psi_1-\Omega^0_1 {d t}  \right)^2 + \ell^2\frac{h_1\mu_2^2}{h_2} \left(d\psi_2-\Omega^0_2dt  \right)^2,
\end{align}
where
\be\label{Omega-0-KK-like}
\Omega^0_1=\Omega^0_2=-\frac{1}{\ell}\sqrt{\frac{q\mathbf{Y}_s}{a^2+q}},\, \qquad \Omega_a^{0R}=\frac{a\Xi_a}{a^2+q}\,
\ee
are the horizon angular velocities at the EVH point, $\mathbf{V}$ is as in \eqref{V0-W} and
\begin{equation}
 h_1^2=\frac{a^2\cos^2\theta+q}{a^2+q}\,, \qquad h_2^2= \frac{a^2\cos^2\theta+q\mu_3^2}{a^2+q}\,.
\label{h1-h2}
\end{equation}

As before, there is a single vanishing cycle responsible for the vanishing entropy and the emergence of a local AdS${}_3$ throat. This corresponds to the isometry direction $\partial_\psi$ along the 3-sphere in the asymptotic AdS${}_5$. Notice that despite having vanishing $J_b, J_3$, rotating EVH black holes still have a single vanishing cycle, keeping the $d\psi_3 + \frac{q}{a\ell}d\psi$ cycle finite. The relevance of this combination is physically understood noticing that
\begin{equation}\label{Omega-xi}
  \Omega_b + \frac{a\ell}{q}\Omega_3 = 0
\end{equation}
(up to ${\cal O}(\epsilon^2)$) in the  near-EVH limit \eqref{delta b-delta M}, where \eqref{ang velocities in ASF}, \eqref{Omega-S5-Phi} and \eqref{eq:5dchepot}
have been used.
Thus, there is {\it no} angular velocity along the $\U(1)$ defined by $d\psi_3 + \frac{q}{a\ell}d\psi$. This point also brings up the question on the uniqueness of the near horizon $\U(1)$ describing the AdS${}_3$ angular momentum. We briefly discuss this matter below.

\paragraph{Choice of IR $\U(1)$ :} The previous discussion suggests to work with
\begin{equation}
  \xi=  \sin\omega_\xi\ \psi_3 + \cos\omega_\xi\ \psi\,,\qquad \tan\omega_\xi=\frac{a\ell}{q}\,.
\label{xi-angle}
\end{equation}
However, there is still some freedom left in the choice of the AdS${}_3$ cycle $\psi$ in \eqref{NH-metric-no-chi-scale}, since it could be replaced by
\be\label{chi-angle}
\chi=  \frac{1}{\cos(\omega_\xi-\omega_\chi)}\left(\cos\omega_\chi\ \psi_3- \sin\omega_\chi\ \psi\right)\,,
\ee
for arbitrary $\omega_\chi$ and as long as $\chi$ does not coincide with $\xi$, i.e. $\omega_\chi-\omega_\xi \neq \pi/2$,\footnote{To be more precise, for our near horizon limit procedure leading to metric \eqref{NH-metric-no-chi-scale} to go through, one needs to assume that $\omega_\chi-\omega_\xi -\frac{\pi}{2}$ is finite (and not of order $\epsilon$).} we will obtain the same near-horizon geometry as in \eqref{NH-metric-no-chi-scale}.
\begin{figure}[t]
 \begin{center}
   \includegraphics[scale=0.55]{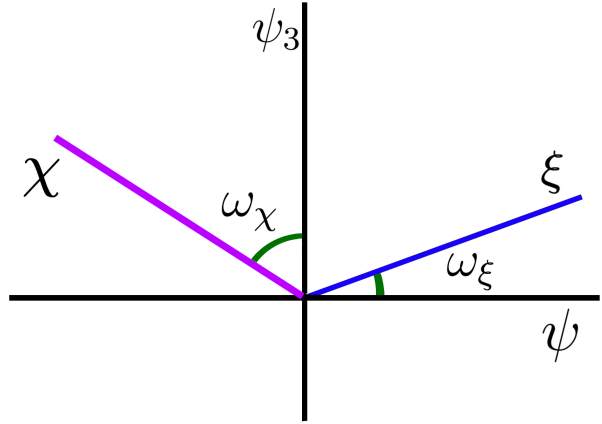}
  \caption{$\chi-\xi$ directions in $\psi_3-\psi$ plane. $\psi$ and $\psi_3$ are respectively coming from \ads{5} and \sph{5} parts of the original geometry.
  $\xi$ is the direction transverse to near horizon \ads{3} while $\chi$ is the angular part of \ads{3}.}
 \label{angles}
 \end{center}
 \end{figure}
The normalization in $\chi$ was chosen so that the volume form in the $\psi_3-\psi$ plane equals the volume form in the $\chi-\xi$ plane, i.e. $d\psi_3\wedge d\psi=d\chi\wedge d\xi$ and both $\chi$ and $\xi$ are ranging over $[0,2\pi]$. In fact, not only the near horizon metric \eqref{NH-metric-no-chi-scale}, but also all physical observables are expected to be independent of $\omega_\chi$. We will see this explicitly in our analysis in the rest of this paper.

Contrary to the static EVH case, which always requires a 10d uplift to have a well defined near horizon geometry, the rotating EVH one does allow a purely 5d description for the particular choice $\omega_\chi=\pi/2$, for which $\chi=\psi$.
In this case, the entire pinching AdS${}_3$ lies inside AdS${}_5$. Its near horizon geometry is presented in Appendix \ref{Appendix-5d-NH-limit}.

After this digression, let us return to the near horizon geometry which is obtained by considering the limit
\begin{align}
\psi_1=\tilde{\psi}_1 + \Omega^0_1\ t  ,&\qquad   \psi_2 = \tilde{\psi}_2+\Omega^0_2\ t,\qquad  \phi=\tilde{\phi} +\Omega_a^{0R}\ t \,, \nonumber \\
t=  \frac{K}{\epsilon}\tau \,,&\qquad   \chi = -\frac{\tilde{\chi}}{\epsilon},\qquad
r=\epsilon\frac{x}{K}\,, \,\,\,\left(K=\sqrt{\frac{\ell^2(a^2+q)}{a^2\ell^2+q^2}}\right)
\label{scalings-NHEVHKK}
\end{align}
on the original black hole metric resulting in
\begin{align}
\label{NHEVHKK}
  ds^2 &=  {h_1h_2}\left[ -\frac{x^2}{\ell_3^2} d\tau^2 +\frac{\ell_3^2dx^2}{x^2}   +x^2 d\tilde{\chi}^2 \right] \\
   &+ \frac{(a^2+q)h_1h_2}{\Delta_\theta}d\theta^2+\frac{\ell^2 \cos^2\alpha \cos^2\theta}{K^2 h_1h_2}d\xi^2 + \frac{a^2+q}{\Xi_a^2} \frac{h_2}{h_1^3}\Delta_\theta \sin^2\theta d\tilde{\phi}^2 \nonumber\\
&+ \ell^2\frac{h_2}{h_1}d\alpha^2 +\ell^2\frac{h_1}{h_2}\sin^2\alpha d\beta^2+
\ell^2\frac{h_1}{h_2} \left[ \mu_1^2 (d\tilde{\psi}_1-Ad\tilde\phi)^2 +\mu_2^2(d\tilde{\psi}_2-Ad\tilde\phi)^2\right] \nonumber,
\end{align}
where $\alpha,\beta,\theta\in[0,\pi/2]$, $\phi\in [0,2\pi]$, $\mu_1=\sin\alpha\cos\beta,\ \mu_2=\sin\alpha\sin\beta$  and
\be\label{A-NHEVHKK}
\Delta_\theta=1-\frac{a^2}{\ell^2}\cos^2\theta,\quad
A=\frac{a \sqrt{ {q\mathbf{Y}_s}} }{\ell \ell_3\Xi_a \sqrt{\mathbf{V}}}\frac{\sin^2\theta}{h_1^2}.
\ee
The first line in \eqref{NHEVHKK} is again conformal to a pinching AdS${}_3$ orbifold with radius
\begin{equation}
 \ell_3^2=\frac{a^2+q}{\mathbf{V}}=\frac{a^2+q}{1+\frac{2q}{\ell^2}+\frac{a^2}{\ell^2}},
\label{ads3-radius-KK-case}
\end{equation}
which reduces to \eqref{AdS3-S3-radii} in the static limit $(a=0)$. This metric solves the type IIB supergravity equations of motion with a constant dilaton and a RR 5-form. The analysis is very similar to that in \cite{Fareghbal:2008ar}.

There are special values in parameter space that deserve special mention. For $q=0$, one recovers Kerr AdS${}_5$, $\xi=\psi_3$ and $\chi$ may be chosen entirely inside AdS$_5$. %\cite{Bardeen:1999px}.
The BPS surface defined by $q=al$ has no special features compared to the generic case \eqref{NHEVHKK}. Finally, the $a\to\ell$ limit (the edge of the EVH plane in figure \ref{EVH}) must be handled with care. Recent work \cite{Berkooz:2012qh} confirms the existence of a 2d chiral spectrum also in this case.

%%%%%%%%%%%%%%%%%%%
\subsection{Rotating near-EVH case}
\label{sec:rot-nearevh}
%%%%%%%%%%%%%%%%%%%

Near-EVH rotating black holes are described by \eqref{delta b-delta M} in parameter space. Their near horizon geometry is obtained through the same limit \eqref{scalings-NHEVHKK} and gives rise to
  \begin{align}
  ds^2 &={h_1h_2}\left[ -\frac{(x^2-x_+^2)(x^2-x_-^2)}{\ell_3^2  x^2} d\tau^2 +\frac{\ell_3^2 x^2dx^2}{(x^2-x_+^2)(x^2-x_-^2)}   +x^2 \left(d\tilde{\chi}- \frac{x_+x_-}{\ell_3 x^2} d{\tau}\right)^2 \right] \nonumber \\
 &+ \frac{(a^2+q)h_1h_2}{\Delta_\theta}d\theta^2+\frac{\ell^2 \cos^2\alpha\cos^2\theta}{K^2 h_1h_2}d\xi^2 + \frac{a^2+q}{\Xi_a^2} \frac{h_2}{h_1^3}\Delta_\theta \sin^2\theta d\tilde{\phi}^2 \label{BTZ-NH-limit final} \\
 & + \frac{\ell^2h_2}{h_1}d\alpha^2 +\frac{\ell^2h_1}{h_2}\sin^2\alpha d\beta^2+
\frac{\ell^2h_1}{h_2} \left[ \mu_1^2 (d\tilde{\psi}_1-Ad\tilde\phi)^2 +\mu_2^2(d\tilde{\psi}_2-Ad\tilde\phi)^2\right] \nonumber,
\end{align}
where  $\mu_1,\ \mu_2,\ h_1,\ h_2,\ \Delta_\theta,\ A$ in \eqref{mu-i}, \eqref{h1-h2} and \eqref{A-NHEVHKK}, and $\alpha,\beta,\theta\in[0,\pi/2]$, $\phi\in [0,2\pi]$.

The first line is conformal to pinching BTZ black holes in the region of the deformation parameter space where the outer and inner horizons $x_\pm$
\be\label{x pm for btz}
x_\pm^2=K^2\frac{r^2_\pm}{\epsilon^2}=\frac{\ell^2(a^2+q)}{q^2+a^2\ell^2}
\left[\frac{\mathbf{W} M\pm\sqrt{\mathbf{W}^2M^2-\mathbf{V}a^2\hb^2}}{\mathbf{V}}\right]\,,
\ee
exist. This holds for\footnote{For $\mathbf{W}^2M < \mathbf{V}a^2\hb^2$ we get a space with a conical defect. This is similar to the situation discussed in \cite{static case}.}
\begin{equation*}
  \mathbf{W}^2 M\geq \mathbf{V}a^2\hb^2.
\end{equation*}
Notice $\hb=0$ corresponds to a vanishing inner horizon, whereas $M=0$ forces $\hb=0$, in this region of parameters. Thus $M$ controls the size of the outer horizon. Notice how the outer and inner horizons of the near-EVH static black hole in \eqref{x-pm-btz-static}  cannot be obtained as a limit of the ones for the rotating case in \eqref{x pm for btz}. As stressed in section \ref{sec:near-evh}, this is due to the fact that the near horizon and the near-EVH limits in the two cases do not commute.

%%%%%%%%%%%%%%%%%%%
\paragraph{Temperature \& entropy:} To use the standard thermodynamic relations satisfied by BTZ black holes, we must compactify \eqref{BTZ-NH-limit final} to three dimensions. This is achieved by considering the ansatz
\begin{align*}
  ds^2 &=  {h_1h_2}g^{(3)}_{\mu\nu} dx^{\mu}dx^{\nu} + \frac{(a^2+q)h_1h_2}{\Delta_\theta}d\theta^2+\frac{\ell^2 \cos^2\alpha\cos^2\theta}{K^2 h_1h_2}d\xi^2 + \frac{a^2+q}{\Xi_a^2} \frac{h_2}{h_1^3}\Delta_\theta \sin^2\theta d\tilde{\phi}^2 \\
&+ \frac{\ell^2h_2}{h_1}d\alpha^2 +\frac{\ell^2h_1}{h_2}\sin^2\alpha d\beta^2+
\frac{\ell^2h_1}{h_2} \left[ \mu_1^2 (d\tilde{\psi}_1-Ad\tilde\phi)^2 +\mu_2^2(d\tilde{\psi}_2-Ad\tilde\phi)^2\right]\,.
\end{align*}
Proceeding as in subsection \ref{sec:nevhstatic}, the 3d Newton's constant is
\begin{equation}
\label{G3-G10-KK-case}
\frac{1}{G_3}=\frac{2N^2\sqrt{(a^2\ell^2+q^2)(a^2+q)} }{\Xi_a\ell^4}\,.
\end{equation}
Note that in the $a=0$ case we recover the static EVH expression \eqref{3d-GN-static}.

The temperature and entropy of the pinching BTZ black holes equal
\be\label{BTZ-temperature-entropy}
\begin{split}
T_{\text{BTZ}}&\equiv\frac{x_+^2-x_-^2}{2\pi x_+\ell_3^2}=\frac{ K}{\epsilon}T_{\text{H}}\,,\\
S_{\text{BTZ}}&\equiv \frac{2\pi \epsilon\cdot x_+}{4G_3}=S_{\text{BH}}\,.
\end{split}
\ee
Once more the 3d entropy matches the original 10d black hole one \eqref{S-T-NH-limit}, while the proportionality of temperatures is consistent with the temporal scaling in \eqref{scalings-NHEVHKK}.

%%%%%%%%%%%%%%%%%%%%
\section{The dual EVH/CFT formulation}
\label{sec:evhcft}
%%%%%%%%%%%%%%%%%%%%

In this section we compute the central charges of the IR 2d CFT that is dual to the asymptotically AdS$_3$ structure emerging in the near horizon of static and rotating EVH black holes, and provide an explicit map between its IR quantum numbers and those of the 4d UV dual to the original black hole. We also discuss how the IR first law of thermodynamics follows from the UV one in \eqref{eq:UVflaw}. We view the latter as further supporting evidence of the EVH/CFT correspondence \cite{KKEVH} reviewed in the introduction.

%%%%%%%%%%%%%%%%%%
\subsection{IR 2d CFT description}
%%%%%%%%%%%%%%%%%%

The central charge $c$ of the IR 2d CFT and the quantum numbers $(L_0,\,\bar L_0)$ describing the gravitational black holes described in the previous section can be extracted from standard AdS${}_3$/CFT${}_2$ \cite{wadia-review} taking the pinching periodicity into proper consideration \cite{Massless-BTZ}
\begin{align}
  c&=\frac{3 \ell_3}{2G_3}\epsilon= \frac{3 (a^2+q)}{\ell^{4}\Xi_a}\sqrt{\frac{a^2\ell^2+q^2}{\mathbf{V}}}\ N^2 \epsilon\,  \label{EVHC} \\
  \ell_3M_{\text{BTZ}} &=L_0 + \bar{L}_0 - \frac{c}{12}=\frac{x_+^2+x_-^2}{8\ell_3G_3}\epsilon\,, \\
  J_{\text{BTZ}} &= L_0 - \bar{L}_0 = \frac{x_+x_-}{4\ell_3G_3}\epsilon\,.
\end{align}
Requiring a {\it finite} central charge to have a {\it finite} gap in this IR 2d CFT is achieved by the {\it large N} limit:
\be\label{N-scaling}
\boxed{N^2\epsilon=\text{fixed}}
\ee
It is manifest the entropy $S_{\text{BH}}$ in \eqref{S-T-NH-limit} remains finite in this limit. It is shown below that the same holds for the excitations $M_{\text{BTZ}},\ J_{\text{BTZ}}$ which will be related with precise combinations of the UV quantum numbers.

\paragraph{Static EVH:} The central charge may be obtained from \eqref{EVHC} at $a=0$
\be\label{central-charge-static}
c_{\text{static}}=\frac{3q^2}{\ell^4\sqrt{1+\frac{2q}{\ell^2}} }\ {N^2\epsilon}\,.
\ee
whereas excitations equal
\begin{equation}\label{MJ-Static-like-case}
\ell_3 M_{\text{BTZ}}	=\frac{2M\mathbf{W}_s-\mathbf{Y}_s(\hat{a}^2+\hat{b}^2)}{4\ell^2\sqrt{\mathbf{V}_s}}\ N^2\epsilon\,, \quad \quad J_{\text{BTZ}}= \frac{\ha\hb}{2\ell^2} N^2\epsilon\,.
\end{equation}
Notice how the static EVH point $(0,0,q;m(q))$ determines the IR 2d CFT by fixing its central charge, whereas its orthogonal deformations \eqref{eq:nevhstatic}-\eqref{near-EVH-static} encode their {\it finite} excitations. Any tangential deformation \eqref{eq:tdef} would have simply changed the value of $q$, which would correspond to a different CFT.
Note also that the BTZ mass has contributions from all three parameters describing the transverse deviations from the EVH surface.

\paragraph{Rotating EVH:} The central charge is given by \eqref{EVHC} with excitations
\be\label{MJ-KK-like}
\ell_3M_{\text{BTZ}} = \frac{\ell_3 K}{2\ell^3\Xi_a }\ M\mathbf{W}\ N^2\epsilon\,,\quad \quad
J_{\text{BTZ}}= \frac{\ell_3 K}{2\ell^3 \Xi_a}\ a\hb \sqrt{\mathbf{V}}\ {N^2\epsilon}
\ee
As before, the rotating EVH point $(a,0,q;m(a,q))$ determines the IR 2d CFT central charge and vacuum structure, whereas its orthogonal deformations \eqref{near-EVH-KK-like}-\eqref{delta b-delta M} encode its excitations.

As we have noted all physical quantities \emph{at} the static EVH point can be recovered from the corresponding  expressions \emph{at} the rotating EVH evaluated at $a=0$ point. This is not, however, true for the excitations above these EVH points. For example, the static values for $M_{\text{BTZ}}$, $J_{\text{BTZ}}$ and $T_{\text{H}}$ cannot be recovered by simply setting $a=0$ (or taking $a\to 0$ limit) of the rotating EVH ones. This is due to the fact that in the static case both $a,b$ scale as $\epsilon$, while in the rotating case, $a$ is finite and  $b\sim\epsilon^2$.

Using Cardy's formula \cite{wadia-review}
\begin{equation}
  S_{\text{CFT}} = 2\pi\sqrt{\frac{c}{6}\left(L_0-\frac{c}{24}\right)} +
 2\pi\sqrt{\frac{{\bar c}}{6}\left(\bar{L}_0-\frac{{\bar c}}{24}\right)}\,,
\label{eq:cardy}
\end{equation}
one can immediately check the bulk entropy \eqref{S-T-NH-limit} is reproduced for both EVH black holes.

%%%%%%%%%%%%%%%%%%
\subsection{EVH/CFT$_2$ vs. AdS$_5$/CFT$_4$}\label{CFT2-CFT4}
%%%%%%%%%%%%%%%%%%

The 10d dimensional black holes \eqref{eq:d10embed} interpolate between asymptotically AdS${}_5$ and locally AdS${}_3$ geometries.  The former has a dual (UV) description in terms of ${\cal N}=4$ SYM, whereas the latter must have a dual (IR) description as advocated above. In this section we relate the quantum numbers of both dual theories.

Since our states are typically non-BPS, the conformal dimension $\Delta_{\text{UV}}$ is expected to ``run'' along the RG flow implemented by the near horizon limit $r=\epsilon\rho$, $\epsilon\to 0$. Hence, its relation to $\Delta_{\text{IR}}$ is non-trivial. On the other hand, given the quantized nature of the remaining conserved $\U(1)$ charges, these will not run.
The analysis of this RG flow in field theory and gravity is beyond the scope of this paper, but we can gain some insight by studying how the quantum numbers of a bulk scalar field probe transform along the flow. Given the isometries of the
original black hole \eqref{5d BH metric}, the UV quantum numbers of this scalar field (in the static AdS${}_5$ frame) will be associated with the eigenvalues of the following operators
\be\label{UV-charges}
 \Delta_{\text{UV}}	=\ell E	=i\ell\partial_t,\;\;
 \quad\quad J_{a,b}		=-i\partial_{\phi^S,\psi^S}\;\;
 \quad\quad J_{i,3}	=-\ell Q_{i,3}	=-i\partial_{\psi_{i,3}}.
  \ee
Similarly, the IR quantum numbers are mapped to
\be\label{UV-charges}
 \Delta_{\text{IR}}	=i\ell_3\partial_\tau,\;\;
 \quad\quad J_{\tchi}=-i\partial_{\tchi}\;\;
\quad\quad J_{\xi}=-i\partial_{\xi}\,.
\ee
From now on, as the notation above suggests, we will identify these eigenvalues with the gravity conserved charges. Though this need not hold generically, it will turn out to provide us with the right intuition.

%%%%%%%%%%%%%%%%%%%%%%%%%%%%%%%%
\subsubsection{IR-UV charge mapping, rotating EVH case}
\label{IR-UV-section.5.2.1}
%%%%%%%%%%%%%%%%%%%%%%%%%%%%%%%%

Given the original 10d black hole charges \eqref{eq:angmom}, \eqref{eq:rcha} and \eqref{eq:mass} and the near-EVH expansion of parameters \eqref{delta b-delta M}, charges fall into two categories in the rotating near-EVH  regime:
\begin{itemize}
\item $Y=\{\Delta, J_1, J_a\}$ having an $\epsilon$ expansion of the form $Y=Y^{(0)}+\epsilon^2 Y^{(2)}$.
\item $Z=\{J_b, J_3\}$ with expansion $Z=\epsilon^2 Z^{(2)}$.
\end{itemize}
$Y^{(0)}$ is the value of charges {\it at the EVH point}, whereas $Y^{(2)},Z^{(2)}$ are the {\it near-EVH excitations}. $\Omega_1$ and $\Omega_a$ have analogous expansions to $J_1$ and $J_a$, with finite $\Omega_1^0$ and $\Omega_a^0$ values at the EVH point \eqref{Omega-0-KK-like} and with $\epsilon^2$ corrections.

As discussed in section \ref{section:NH-KK-EVH}, the $J_b$ and $J_3$ charges
\be\label{Jb1 and J31}
 J_b	=\frac{\hb(q^2+a^2\ell^2+q\ell^2)N^2}{2\ell^5\Xi_a}\epsilon^2,\;\;
 \quad\quad J_3=\frac{a\hb qN^2}{2\ell^4\Xi_a}\epsilon^2\,,
\ee
are naturally encoded in the IR geometry in terms of
\begin{align}
 J_{{\tchi}} &=-\frac{1}{\epsilon}\left(-\sin\omega_\xi J_b+\cos\omega_\xi J_3\right)=
\frac{(a^2+q)}{2\ell^2\Xi_a\sqrt{a^2\ell^2+q^2}}a\hb\ N^2\epsilon = J_{{\text{BTZ}}}\,,\label{Jchi} \\
 J_\xi & = \frac{1}{\cos(\omega_\xi-\omega_\chi)}\left(\sin\omega_\chi J_3+\cos\omega_\chi J_b\right)\sim     N^2\epsilon^2\,,
\end{align}
where we used
\be\label{psi-chi-xi}
\psi_3=\cos\omega_\xi \chi+\frac{\sin\omega_\chi}{\cos(\omega_\chi-\omega_\xi)}\xi\,,\quad \psi=-\sin\omega_\xi \chi+\frac{\cos\omega_\chi}{\cos(\omega_\chi-\omega_\xi)}\xi\,.
\ee
The AdS${}_3$ pinching is responsible for the $1/\epsilon$ factor in the first equality of \eqref{Jchi}. This allows $J_{\tilde{\chi}}$ to remain finite, matching the BTZ angular momentum, whereas $J_\xi\sim \epsilon$ is subleading, in the limit \eqref{N-scaling}. This prevents any unphysical dependence on the mixing angle $\omega_\chi$ to survive our limit, as expected on physical grounds.

Let us consider the IR conformal dimension $\Delta_{\text{IR}}$. If we proceed analogously to the other charges, we learn
\begin{align}
 \Delta_{\text{IR}}\equiv i\ell_3 \frac{\partial}{\partial \tau} &=
\frac{\ell_3}{\ell}\frac{K}{\epsilon}\left(i\ell\frac{\partial}{\partial t}+i\ell\Omega_a^{0S}\frac{\partial}{\partial \phi}
+\sum_{i=1,2}i\ell\Omega_i^0\frac{\partial}{\partial \psi_i}\right)\cr
  &=\frac{\ell_3}{\ell}\frac{K}{\epsilon}\left(\Delta-\ell\Omega_a^{0S}J_a
-2\ell\Omega_1^0J_1\right)\label{DeltaIR-KK}.
\end{align}
Using the 5d charges \eqref{eq:rcha}, \eqref{eq:angmom} and \eqref{eq:mass} in the near-EVH regime \eqref{delta b-delta M}, we find
\begin{align}\label{M-BTZ-DeltaIR-KK}
\Delta_{\text{IR}} &=\frac{K\ell_3}{\ell\epsilon}(\Delta^{(0)}-\ell\ \Omega_a^{0S}J_a^{(0)}-2\ell\ \Omega_1^0J_1^{(0)}) +\frac{K\ell_3}{\ell}\epsilon \left(\Delta^{(2)}-\ell\ \Omega_a^{0S}J_a^{(2)}-2\ell\ \Omega_1^0J_1^{(2)}\right)\nonumber \\
&=\Delta^0_{\text{IR}}+\ell_3 M_{\text{BTZ}}\,,
\end{align}
where we used \eqref{MJ-KK-like}, the identity $\ell M_{\text{BTZ}}=K(\Delta^{(2)}-\ell\Omega_a^0J_a^{(2)}-2\ell\Omega_1^0J_1^{(2)})\epsilon$ and $\Delta^0_{\text{IR}}$ is defined as
\begin{equation}\label{zero-point-energy-KK}
\Delta^0_{\text{IR}}=\frac{\ell_3}{\ell}\frac{K}{\epsilon}\frac{N^2(a^2\ell^2-q^2)}{4\ell^4\Xi_a}\,.
\end{equation}
Notice $\Delta^0_{\text{IR}}$ only depends on the rotating EVH point (and not the excitations) and could consequently be interpreted as a ``zero point energy'' from the IR 2d CFT perspective. This contribution is generically divergent, but vanishes when supersymmetry is preserved, i.e. $q=a\ell$. This would reproduce the expected $\Delta_{\text{IR}} = L_0+\bar L_0$ bound in this case, due to the protection of supersymmetry along the RG flow. Near the BPS point, i.e. $q=a \ell-\epsilon^2 \delta_s^2/(2a\ell)$, the 'zero point energy' still remains finite.
\begin{equation}
 \Delta_{\text{IR}}=\frac{\ell_3}{\ell}\frac{K N^2\epsilon}{4\ell^4\Xi_a}\delta_s^2+\ell_3M_{\text{BTZ}}.
\end{equation}

We conclude that both $\Delta_{\text{IR}}$ and $J_{\tchi}$ match the expected IR 2d CFT quantities. This is a good piece of evidence in favour of the EVH/CFT correspondence since despite lack of supersymmetry, the RG flow from the UV to the deep IR does not lead to contributions not captured by the 2d CFT. This fact supports the expectation that the near-EVH sector in the UV dual 4d CFT is a decoupled sector described by this IR 2d CFT where the UV quantum numbers were reshuffled as explained above.

%%%%%%%%%%%%%%%%%%%%%%%%%%%
\subsubsection{IR-UV mapping, static EVH case}
\label{IR-UV-section.5.2.2}
%%%%%%%%%%%%%%%%%%%%%%%%%%%

Proceeding as in section \ref{IR-UV-section.5.2.1}, we distinguish three categories of 10d black holes charges in the
near-EVH static regime \eqref{near-EVH-static}:
\begin{itemize}
\item $Y=\{\Delta, J_1, J_2\}$ having an $\epsilon$ expansion of the form $Y=Y^{(0)}+\epsilon^2 Y^{(2)}$.
\item $X=\{J_a,J_b\}$ having an $\epsilon$ expansion of the form $X=\epsilon X^{(1)}$.
\item $J_3$ with expansion $J_3=\epsilon^2 J_3^{(2)}$.
\end{itemize}
Since all charges scale like $N^2$, charges $X$ have a finite value due to \eqref{N-scaling}.

Given the large gauge transformations \eqref{coord-scaling-static-EVH} and \eqref{phi-psi-NEVH-static-scaling} implemented in the near-EVH static regime, we learn the angular momentum along the AdS${}_3$ pinching circle equals
\be\label{J-chi-static-like}
\begin{split}
J_{\tchi}&=-i\partial_{\tchi}=-i\left(-\frac{1}{\epsilon}\partial_{\psi_3}+\frac{\hb \ell}{q}\partial_\phi+\frac{\ha \ell}{q}\partial_\psi\right)=-\frac{1}{\epsilon}J_3+\frac{\ell}{q}(\ha J_b+\hb J_a)\\
&=J_{\text{BTZ}}+\frac{\ell}{2q}(\ha J_b+\hb J_a),
\end{split}
\ee
where we used \eqref{MJ-Static-like-case}. Furthermore, from \eqref{eq:angmom} and \eqref{near-EVH-static}
$$\frac{\ell}{2q}(\ha J_b+\hb J_a)=\frac{\ha \hb \mathbf{Y}_s}{2\ell^2}\ N^2\epsilon.$$
The IR conformal dimension equals
\be\label{Delta-IR-static-like-1}
\begin{split}
\Delta_{\text{IR}}&=i\ell_3\partial_{\tau}=
\frac{\ell_3}{\sqrt{q}\epsilon}\left[\Delta_{\text{UV}}-2\ell\ \Omega^0_1J_1-\frac{\ell \mathbf{Y}_s}{q}(b J_b+a J_a)\right]\\
&=\Delta^0_{\text{IR}}+\frac{\ell_3\epsilon}{\sqrt{q}}\left(\Delta^{(2)}_{\text{UV}}-
2\ell\ \Omega^0_1J_1^{(2)}-\frac{\ell  \mathbf{Y}_s}{q}(\ha J_a^{(1)}+\hb J_b^{(1)})\right)\,,
\end{split}
\ee
where the ``zero point energy'' $\Delta^0_{\text{IR}}$ is defined as
\be
\Delta^0_{\text{IR}}=\frac{\ell_3}{\sqrt{q}\epsilon}(\Delta_{\text{UV}}^{(0)}-2\ell\ \Omega^0_1J_1^{(0)})=-\frac{\ell_3}{\sqrt{q}\epsilon}\frac{N^2q^2}{4\ell^4}\,.
\ee
Note the zero point energy of the static EVH case $\Delta^0_{\text{IR}}$ can be obtained as the $a=0$ limit of the rotating EVH case \eqref{zero-point-energy-KK}. In the BPS case where $q\sim \epsilon^{1/2}$ \cite{Fareghbal:2008ar,static case} $\Delta^0_{\text{IR}}$ remains finite\footnote{We comment that the negative value of $\Delta^0_{\text{IR}}$ from the dual 2d CFT viewpoint may be attributed to the $-c/12$ Casimir energy of the theory on the cylinder.}. Using the expressions for mass and angular momenta \eqref{eq:angmom}, \eqref{eq:rcha} and \eqref{eq:mass} and the near-EVH scalings \eqref{near-EVH-static} we find
\be\label{DIR-static-middle-stage}
\begin{split}
\Delta_{UV}^{(2)}-2\ell\ \Omega^0_1J_1^{(2)}&=\frac{N^2}{2\ell^2}\left[M  \mathbf{W}_s+\frac{q \mathbf{Y}_s}{2\ell^2}(\frac{\ha^2+\hb^2}{\ell^2})\right]\,,\\
\frac{\ell}{q}(\ha J_a+\hb J_b)&=\frac{(\ha^2+ \hb^2) \mathbf{Y}_s }{2\ell^2}{N^2\epsilon}\,,
\end{split}
\ee
and hence
\be\label{Delta-IR-static-like}
\Delta_{\text{IR}}=\Delta^0_{\text{IR}}+ \ell_3 M_{\text{BTZ}}-\frac{\ell_3}{\sqrt{q}}\frac{\ell}{2q}(\ha J_a+\hb J_b)\,,
\ee
where $M_{\text{BTZ}}$ is given in \eqref{MJ-Static-like-case}.

Thus, even after removing the 'zero point energy', both $J_{\tchi}$ and $\Delta_{\text{IR}}$ contain extra terms compared with the expected BTZ mass $M_{\text{BTZ}}$ and angular momentum $J_{\text{BTZ}}$\footnote{As our ``chain rule'' analysis indicates, it is possible to define new $\tilde{\phi}$ and $\tilde{\psi}$ coordinates such that \eqref{Delta-IR-static-like} and \eqref{J-chi-static-like} respectively reduce to simply $\Delta_{\text{IR}}=\Delta^0_{\text{IR}}+ \ell_3 M_{\text{BTZ}}$ and $J_{\tchi}=J_{\text{BTZ}}$. The new $\tilde\phi$ and $\tilde\psi$ which do this are
$$
\tilde{\psi}={\psi}-\frac{\hb }{2q}\frac{\ell}{\sqrt{q}}\tau\  - \frac{\ha \ell}{{2q}} \tilde{\chi}\,,\quad  \tilde{\phi}={\phi}-\frac{\ha}{2q}\frac{\ell}{\sqrt{q}} \tau - \frac{\hb \ell}{{2q}}  \tilde{\chi}\,.
$$
}. As mentioned in the discussion section, these terms should be associated with some notion of spectral flow in the dual CFT. Nevertheless, we will show below that the contribution of these terms to the IR first law cancels, as expected under spectral flow.

%%%%%%%%%%%%%%%%%%%%%%%%%%%%%%%
\subsection{First law of thermodynamics, IR vs. UV, 3d vs. 5d}
%%%%%%%%%%%%%%%%%%%%%%%%%%%%%%%

Having identified the UV-IR charge map, we now derive the IR first laws from the UV one in \eqref{eq:UVflaw}.
Physical variations appearing in the first law are generically defined as one-forms on the black hole parameter space. In our examples,  the UV $\{dE, dJ_a, dJ_b, dJ_i, dJ_3\}$ forms are defined on a four dimensional space spanned by $(a,b,q,m)$, while in the IR, physical variations belong to the subspace of orthogonal deformations to the EVH hyperplane, leaving the EVH point fixed. The latter are computed using the near-EVH charge expansions worked out in sections \ref{IR-UV-section.5.2.1} and \ref{IR-UV-section.5.2.2} inserting them into \eqref{eq:UVflaw}. Below, we consider static and rotating EVH cases separately.

\subsubsection{Rotating EVH case}

Consider the UV first law \eqref{eq:UVflaw}. Using \eqref{BTZ-temperature-entropy}, its left hand side equals
\be
T_{\text{H}} dS_{\text{BH}} = \frac{\epsilon}{K}T_{\text{BTZ}} dS_{\text{BTZ}} \,.
\ee
To discuss its right-hand-side, notice first that the maps \eqref{DeltaIR-KK}-\eqref{M-BTZ-DeltaIR-KK} yield
\be\label{dM-BTZ-KK}
\begin{split}
dM_{\text{BTZ}}&=\frac{1}{\ell_3} d\Delta_{\text{IR}}={K}{\epsilon}\left({dE^{(2)}}-2\Omega_1^0dJ_1^{(2)}-\Omega_a^{0R}dJ_a^{(2)}\right) \\
&=\frac{K}{\epsilon}\left({dE}-2\Omega_1dJ_1-\Omega_a^{R}dJ_a\right)+{\cal O}(\epsilon^2).
\end{split}
\ee
Notice we explicitly dropped the piece coming from the variation of the `zero point energy' $d\Delta_{\text{IR}}^0$ since the latter is {\it independent} of the near-EVH excitations. Second, using
\be\begin{split}
\Omega^0_b=\frac{a}{q+a^2}\frac{ab}{r_+^2}\,&,\qquad  \Omega^0_3=-\frac{q}{l(q+a^2)}\frac{ab}{r_+^2}\,,\\
\Omega_b^0+\frac{la}{q}\Omega_3^0=0\,,&\qquad \tan\omega_\xi=\frac{a\ell}{q}\,,
\end{split}
\ee
and \eqref{Jchi}, we derive
\be\label{Omega-J-BTZ}
\begin{split}
\Omega_b dJ_b+\Omega_3dJ_3&=\epsilon^2(\Omega_b^0dJ_b^{(2)}+\Omega_3^0dJ_3^{(2)})+{\cal O}(\epsilon^3)\\
&=\frac{\epsilon}{K} \Omega_{\text{BTZ}}dJ_{\text{BTZ}}\,,
\end{split}
\ee
where we used
\be\label{omega-chi-xi}
\Omega_{\text{BTZ}}\equiv\frac{x_+x_-}{x_+^2\ell_3}=\frac{K}{\epsilon}\Omega_\chi=\frac{K}{\epsilon}(\cos\omega_\xi \Omega^0_3-\sin\omega_\xi \Omega^0_b)\,.
\ee
Note that as expected, the mixing angle $\omega_\chi$ does not appear in any of the expressions above.

Adding all contributions and dropping the overall $\epsilon/K$ constant factor, the exact IR first law is derived
\begin{equation}
\boxed{\begin{split}
T_{\text{H}}dS_{\text{BH}}&=dE-2\Omega_1 dJ_1-\Omega_a dJ_a-\Omega_b dJ_b-\Omega_3 dJ_3\quad \\
& \Downarrow \\
\quad T_{\text{BTZ}}dS_{\text{BTZ}}&=dM_{\text{BTZ}}-\Omega_{\text{BTZ}}dJ_{\text{BTZ}}
\end{split}}
\end{equation}
where we already dropped all vanishing subleading contributions in the $\epsilon\to 0$ limit.
Notice the $\xi$ direction does not contribute above because the $\Omega_\xi dJ_\xi$ term is an order $\epsilon^2$ smaller than the leading term.

\subsubsection{Static EVH case}
\label{sec:staticf}

Using \eqref{BTZ-temperature-entropy-static}, the left hand side of the UV first law scales like
\begin{equation}
  T_{\text{H}}dS_{\text{BH}} = \epsilon\frac{\sqrt{q}}{\ell}T_{\text{BTZ}}dS_{\text{BTZ}}.
\end{equation}
When analysing the right hand side, recalling \eqref{omega-0-1}, \eqref{Delta-IR-static-like-1}, \eqref{DIR-static-middle-stage} and \eqref{Delta-IR-static-like} we have
\begin{equation}\label{dE-OmegaJ-static}
  dE-\sum_{i=1,2}\Omega_i^0dJ_i =  \frac{\sqrt{q}\epsilon}{\ell}\left(dM_{\text{BTZ}}
+\frac{\ell  \mathbf{Y}_s}{2q^{3/2}}\ d\left(\ha J_a + \hb J_b\right)\right).
\end{equation}
Since the angular velocities equal (recall \eqref{ang velocities in ASF} and \eqref{Omega-S5-Phi})
\begin{align}\label{psi 3 ang velocity near static}
\Omega_{BTZ} &\equiv\frac{x_-x_+}{\ell_3 x_+^2}=\frac{\ha\hb}{\sqrt{ \mathbf{V}_s}}\frac{1}{\ell_3 \hat{r}_+^2}=-\frac{\ell}{\sqrt{q}}\Omega_{3}+{\cal O}(\epsilon^2)\,\cr
\Omega_a & =\frac{a(b^2+r_+^2\mathbf{Y}_s)}{ q r_+^2}=\left(\frac{\hb}{\sqrt{q}}\Omega_{\text{BTZ}}+\frac{\ha  \mathbf{Y}_s }{q}\right)\epsilon+{\cal O}(\epsilon^3)\,, \\
\Omega_b &=\frac{b(a^2+r_+^2\mathbf{Y}_s)}{qr_+^2}=
\left(\frac{\ha}{\sqrt{q}}\Omega_{\text{BTZ}}+\frac{\hb  \mathbf{Y}_s}{q}\right)\epsilon+{\cal O}(\epsilon^3)\,,\nn
\end{align}
we derive
\begin{align}
  \Omega_adJ_a+\Omega_bdJ_b &= \frac{\sqrt{q} \mathbf{Y}_s \epsilon}{\ell}\left(\Omega_{\text{BTZ}}dJ_{\text{BTZ}}+\frac{\ell}{2q^{3/2}} d\left(\ha J_a + \hb J_b\right)\right),\label{JJ-static} \\
 \Omega_3dJ_3 &=\frac{\sqrt{q}\epsilon}{\ell}\left(-\frac{q}{\ell^2}\Omega_{\text{BTZ}}dJ_{\text{BTZ}}\right).
\end{align}

Adding all terms, we notice all the $\Omega_{\text{BTZ}}$ dependence arranges itself as $-{\epsilon\sqrt{q}/\ell}\Omega_{\text{BTZ}}dJ_{\text{BTZ}}$, whereas all dependence on $dJ_a$ and $dJ_b$ cancels, so that, modulo the overall ${\epsilon\sqrt{q}/\ell}$ factor
\begin{equation}\label{IR-first-law-static}
\boxed{\begin{split}
T_{\text{H}}dS_{\text{BH}}&=dE-2\Omega_1 dJ_1-\Omega_a dJ_a-\Omega_b dJ_b-\Omega_3 dJ_3 \\
& \Downarrow \\
T_{\text{BTZ}}dS_{\text{BTZ}}&=dM_{\text{BTZ}}-\Omega_{\text{BTZ}}dJ_{\text{BTZ}}
\end{split}}
\end{equation}
where we already dropped all vanishing subleading contributions in the $\epsilon\to 0$ limit. Notice, in particular, how the   extra dependence on $J_a$ and $J_b$ in \eqref{Delta-IR-static-like} and \eqref{J-chi-static-like} drops from the IR first law. This highlights the invariance of the first law under large gauge transformations generating constant electric and magnetic terms on the transverse 3-sphere. Such invariance occurs in the standard D1-D5 dual CFT and we expect something similar to hold here.

%%%%%%%%%%%%%%%%%%
\section{Relation between EVH/CFT and Kerr/CFT}
\label{sec:kerrcft}
%%%%%%%%%%%%%%%%%%

In this section we seek a connection between the 2d CFTs appearing in the EVH/CFT correspondence discussed in previous sections and the 2d chiral CFTs emerging in the Kerr/CFT correspondence \cite{Kerr-CFT}. Our perspective is that a 2d chiral CFT is nothing but the Discrete Light-Cone Quantization (DLCQ) of a standard 2d CFT
\cite{Balasubramanian:2009bg}, as was already discussed in \cite{static case} for static charged \ads{5} EVH black holes.
To explore this perspective, we first review the Kerr/CFT formalism and apply it to the family of black holes \eqref{5d BH metric} and their 10d embeddings \eqref{eq:d10embed}. In the region of parameters where Kerr/CFT and EVH/CFT overlap, we can always derive one of the Kerr/CFT descriptions from the EVH/CFT. Some of the caveats that require a better understanding are mentioned in the discussion section \ref{discussion-section}.

\paragraph{Review of Kerr/CFT for \ads{5} black holes \cite{Chow:2008dp}.} The Kerr/CFT correspondence applies to any extremal black hole of finite horizon size. In our notation, one considers the near horizon limit
\begin{equation}
r=r_0(1+\epsilon y)\,,\qquad t=\frac{\tau}{2\pi {T_{\text{H}'}}^0r_0\epsilon}\,,\qquad \hat{\phi}=\phi+{\Omega_a}^0t\,,\qquad \hat{\psi}=\psi+{\Omega_b}^0t\,,
\label{eq:nhregular}
\end{equation}
where $r_0$ stands for the extremal horizon, $T_{\text H}'$ is the derivative of Hawking temperature w.r.t. the horizon size $r_0$ and all 0 indices refer to the thermodynamical quantities being evaluated at it. Using the Taylor expansion for the function $X$ in \eqref{5d BH metric} controlling the horizon size,
\begin{equation}
X=(r-r_0)^2\frac{X''(r_0)}{2}+\mathcal{O}(\epsilon^3)\equiv V_f(r-r_0)^2, \,\,{\rm with} \,\, V_f=1+\frac{3a^2b^2}{r_0^4}+\frac{6r_0^2+a^2+b^2+2q}{\ell^2}\,,
\end{equation}
the resulting metric describes an S${}^3$ bundle over AdS${}_2$ \cite{Chow:2008dp}
\begin{equation*}\label{5d-NH-ext}
ds_5^2=A(\theta)\left(-y^2d\tau^2+\frac{dy^2}{y^2}\right)+F(\theta)d\theta^2+B_1(\theta)e_1^2+B_2(\theta)\left(e_2+C(\theta)e_1\right)^2\,,
\end{equation*}
where the scalar functions are given by
\begin{align*}
A(\theta)=\frac{X_1^0(\rho_0^2+q)}{V_f}\,&,\qquad F(\theta)=\frac{X_1^0(\rho_0^2+q)}{\Delta},	\\
B_1(\theta)=X_1^0\left(g^0_{\phi\phi}-\frac{{g^0_{\phi\psi}}^2}{g^0_{\psi\psi}}\right)&\,,\quad	
B_2(\theta)=X_1^0g^0_{\psi\psi}\,,\quad C(\theta)=\frac{g^0_{\phi\psi}}{g^0_{\psi\psi}}\,,
\end{align*}
\begin{eqnarray*}
g^0_{\phi\phi}=\frac{(Z+b^2C\sin^2\theta)\sin^2\theta}{(\rho_0^2+q)f_2^2\Xi_{a}^2}\Bigg|_{\epsilon\to 0}\,,\
g^0_{\psi\psi}=\frac{(W+a^2C\cos^2\theta)\cos^2\theta}{(\rho_0^2+q)f_1^2\Xi_{b}^2}\Bigg|_{\epsilon\to 0}\,,\
g^0_{\phi\psi}=\frac{abC\sin^2\theta\cos^2\theta}{(\rho_0^2+q)f_1f_2\Xi_{a}^2\Xi_{b}^2}\Bigg|_{\epsilon\to 0}\,,
\end{eqnarray*}
whereas the set of one-forms $e_{a}=d\phi_a+k_{\phi_a}yd\tau$ for $a=1,2$ (in terms of our earlier notation $\phi_1=\phi$ and $\phi_2=\psi$) is determined by the constants
\begin{equation}
k_{\phi}=\frac{2a\Xi_{a}({f_2^0}^2+b^2q)}{f_3^0r_0V_f}		\,,\qquad	 k_{\psi}=\frac{2b\Xi_{b}({f_1^0}^2+a^2q)}{f_3^0r_0V_f}\,.
 \label{5d ka, kb}
\end{equation}
According to the Kerr/CFT dictionary reviewed in \cite{Kerr-CFT-review}, these fix the dual CFT Frolov-Thorne temperatures
\begin{equation}
  T_{\phi_a}=\frac{1}{2\pi k_{\phi_a}}\,,
\end{equation}
whereas the central charges of the chiral Virasoro algebra obtained from the asymptotic symmetry group analysis equal
\cite{Chow:2008dp}
\be%
c_{\phi}=\frac{6\pi a[(r_0^2+b^2)^2+qb^2]}{G_5V_f\Xi_{b}r_0^2}		\,,\qquad	c_{\psi}=\frac{6\pi b[(r_0^2+a^2)^2+qa^2]}{G_5V_f\Xi_{a}r_0^2}.	
\label{5d ca, cb}
\ee%

\paragraph{Embedding to 10 dimensions:} As originally discussed in \cite{Kerr-CFT}, when the 5d geometry is embedded in higher dimensions, as in \eqref{eq:d10embed}, the number of geometrical $\U(1)$ isometries that can get enhanced to a full Virasoro is enlarged. Proceeding as before, we write the 10d near horizon geometry as
\begin{align*}
 ds_{10}^2	 &=\tilde{A}(\theta_n)\left(-y^2d\tau^2+\frac{dy^2}{y^2}\right)+\tilde{B_1}(\theta_n){e_{\phi}}^2+\tilde{B_2}(\theta_n)\left(e_{\psi}+C(\theta_0)^2e_{\phi}\right)^2\\
	 &+\sum_{n,m=0}^2F_{\theta_n\theta_m}(\theta_n,\theta_m)\ d\theta_nd\theta_n+\sum_{i=1}^3D_i(\theta_n)\left(e_{\psi_i}+P_i(\theta_0)(e_{\phi}+e_{\psi})\right)^2,
\end{align*}
with $\theta_0$ being the latitudinal coordinate in AdS${}_5$ and $\theta_1,\theta_2$ those of the transverse S$^5$ (the same as $\alpha, \beta$ angles defined in \eqref{mu-i}), and
\be%
\tilde{A}(\theta_n)=\sqrt{\tilde{\Delta}}A(\theta_0)	\,,\qquad	 \tilde{B}_{1,2}(\theta_n)=\sqrt{\tilde{\Delta}}B_{1,2}(\theta_0)	\,,\qquad D_i(\theta_n)=\frac{\mu_i^2}{X_i^0}\nn.
\ee%
where $\tilde\Delta$ is defined in \eqref{Delta-tilde}.
The non-zero $F_{\theta_m\theta_n}$ are
\begin{align*}
F_{\theta_0\theta_0}=\sqrt{\tilde{\Delta}}F(\theta)\,,\qquad 	
F_{\theta_1\theta_1}=H_0&(\cos^2\theta_1+\sin^2\theta_1\sin^2\theta_2)+\sin^2\theta_1\cos^2\theta_2\,,	\\
F_{\theta_2\theta_2}=H_0(\cos^2\theta_1\cos^2\theta_2)+\cos^2\theta_1\sin^2\theta_2\,,\quad
&	F_{\theta_1\theta_2}=\sin\theta_1\sin\theta_2\cos\theta_1\cos\theta_2(1-H_0).\nn
\end{align*}
This metric can be viewed as a warped S${}^3\times$S${}^5$ bundle over AdS${}_2$. The corresponding Frolov-Thorne temperatures are fixed by
 \begin{align}
  k_{\psi_1}= k_{\psi_2}	&=\frac{2r_0^3q\sqrt{1+2m/q}}{\ell^3f_3^0V}(2r_0^2+a^2+b^2+2q),\nn			\\
 k_{\psi_3}	&=-\frac{2  abq}{\ell r_0f_3^0V_f}(2r_0^2+a^2+b^2+q)\,,	\label{10d ki}	
 \end{align}
corresponding to the three $\U(1)$s in the 5-sphere, whereas the central charges of the corresponding CFTs are
\footnote{Negative central charge $c_{\psi_3}$ may sound alarming. However, we note that in a 2d CFT the sign which has physical
significance is $c L_0$ or $\frac{1}{c}L_0$ and that the charge corresponding to rotations on $\psi_3$,
$Q_3$ is negative in our conventions \eqref{eq:rcha}; had we chosen the opposite orientation for $\psi_3$, both $c_{\psi_3}$
and $Q_3$ would have changed sign. \label{negative-c}}
\begin{equation}
c_{\psi_1}= c_{\psi_2}=\frac{6\pi r_0^2 q\sqrt{1+2m/q} (2r_0^2+a^2+b^2+2q)}{\ell^3 G_5\Xi_{a}\Xi_{b}V_f}\,,\qquad	
c_{\psi_3}=-\frac{6\pi abq(2r_0^2+a^2+b^2+q)}{\ell G_5\Xi_{a}\Xi_{b}V_fr_0^2}\,.	
\label{10d ci}
\end{equation}

Kerr/CFT suggests the existence of five apparently inequivalent chiral CFTs reproducing the entropy of the extremal black holes upon using Cardy formula
\begin{equation}
  S = \frac{\pi^2}{3}c_i T_i
\end{equation}
where $T_i$ is the corresponding Frolov-Thorne temperature \cite{Chow:2008dp}\footnote{For $\psi_3$ direction which the central charge was negative, one may directly show that the Frolov-Thorne temperature is also negative, \emph{cf}. discussion in footnote \ref{negative-c}.}
\be
T_i=-\frac{\partial T_{\text{H}}/\partial r_+}{\partial\Omega_i/\partial r_+}\bigg|_{r_+=r_0}\,.
\ee
It may be surprising that a given black hole has many dual descriptions. At the classical level, it was shown in \cite{Loran} that there exist some transformations, leaving the near horizon metric invariant, relating different CFTs in an infinite lattice of them.

\paragraph{Taking the near-EVH limit:} Since Kerr/CFT works for extremal finite size black holes,
while EVH/CFT works for near-EVH black holes which can be extremal or non-extremal, we need to compare
them in some region of parameter space where both apply. This can be achieved by restricting to the extremal excitations in the EVH/CFT side, i.e. when the BTZ geometry obtained in the near horizon limit of near-EVH black holes is an extremal BTZ, and considering the vanishing entropy limit in the Kerr/CFT side. The second step involves
a singular limit. On the CFT side, this is because some of the Kerr/CFT central charges tend to zero trying to reproduce the appearance of a vanishing geometric cycle to account for the vanishing entropy. On the bulk side, this is because of the non-commutativity between taking the near horizon limit of a near-EVH black hole and taking the near-EVH limit of the near horizon of an extremal finite horizon black hole. The two limit procedures lead to different geometries.

Despite these concerns, we will show the Kerr/CFT central charge associated with the vanishing $\U(1)$ isometry cycle remains finite in the EVH limit and always matches the standard AdS${}_3$ Brown-Henneaux central charge computed in the EVH/CFT correspondence.

\paragraph{Rotating EVH case.}
The leading terms in the Kerr/CFT central charges in the EVH limit \eqref{delta b-delta M}  take the form

\begin{align}\label{Kerr/CFT-central-charges-rotating}
c_\phi=\frac{3\hb}{\ell}\frac{q+a^2{\mathbf{V}^{-1}}}{\ell^2\sqrt{\mathbf{V}}} N^2\epsilon^2\,&,\qquad c_{\psi_1}=c_{\psi_2}=\frac{3\sqrt{q}}{\ell}\frac{a\hb}{\ell^2\mathbf{V}}\frac{\ell_3}{\ell}\sqrt{\mathbf{Y}_s} N^2\epsilon^2\,,\\
c_\psi=\frac{3a\sqrt{\mathbf{V}}}{\ell\Xi_a}\frac{\ell_3^2}{\ell^2} N^2\,
&,\qquad c_{\psi_3}=-\frac{3q\sqrt{\mathbf{V}}}{\ell^2\Xi_a}\frac{\ell_3^2}{\ell^2} N^2\,,
\end{align}
where we used the identities $V_f=4\mathbf{V}$ and $r_0^2=ab/\sqrt{\mathbf{V}}$. In the infinite $N$ limit
\eqref{N-scaling}, $c_\phi,\ c_{\psi_1},\ c_{\psi_2}\sim \epsilon \to 0$. Thus, the corresponding CFTs break down.
Conversely, $c_\psi$ and $c_{\psi_3}$ diverge. Nonetheless, we already discussed in section  \ref{sec:nh} that the
relevant IR $\U(1)$s are given by \eqref{psi-chi-xi}. Following \cite{Loran}, the central charges transform like
\be\label{c-chi-c-xi}
c_\xi=\cos\omega_\xi c_\psi+\sin\omega_\xi c_{\psi_3}\,,\qquad c_{\tchi}=-\frac{\epsilon}{\cos(\omega_\xi-\omega_\chi)}\left(\cos\omega_\chi c_{\psi_3}-\sin\omega_\chi c_{\psi}\right)\,,
\ee
under the coordinate transformations \eqref{psi-chi-xi}. These equal
\be\label{c-xi-c-chi-values}
c_\xi=0+{\cal O}(\epsilon)\,,\qquad c_{\tchi}=\frac{3\sqrt{\mathbf{V}}}{\ell^2\Xi_a}\frac{\ell_3^2}{\ell^2}\frac{\sqrt{q^2+a^2\ell^2}}{\ell^2}\ N^2\epsilon\,,
\ee
where we used $\tan\omega_\xi=a\ell/q$. The vanishing of $c_\xi$ agrees with the absence of angular velocity and
momentum. More importantly, $c_{\tchi}$ exactly equals the Brown-Henneaux central charge in \eqref{EVHC}.
This latter is in accord with our proposal/vision for connecting Kerr/CFT and EVH/CFT:
the chiral 2d CFT appearing in Kerr/CFT is the DLCQ of the one appearing in the EVH/CFT.

\paragraph{Static EVH regime.}
The leading terms in the Kerr/CFT central charges in the EVH limit \eqref{near-EVH-static} take the form
\begin{align}\label{Kerr/CFT-central-charges-static}
c_\phi=\frac{3q}{\ell^2\sqrt{\mathbf{V}_s}} \frac{\hb}{\ell} N^2\epsilon\,&,\qquad c_\psi=\frac{3q}{\ell^2\sqrt{\mathbf{V}_s}} \frac{\ha}{\ell} N^2\epsilon\\
c_{\psi_1}=c_{\psi_2}=\frac{6\sqrt{q} \ell_3^3}{\ell^4}\frac{\ha\hb}{\ell^2}\sqrt{ \mathbf{Y}_s} N^2\epsilon^2\,&,\qquad c_{\psi_3}=-\frac{3q^2}{\ell^4\sqrt{\mathbf{V}_s}} N^2\,,
\end{align}
where we used the identities $V_f=4\mathbf{V}_s$ and $r_0^2=ab/\sqrt{\mathbf{V}_s}$. As in the rotating EVH case,
we are interested in identifying the central charges for the relevant IR $\U(1)$s. Following \cite{Loran},
 \eqref{coord-scaling-static-EVH} implies
\be\label{c-ker-cft-c-static}
c_{\hat{\chi}}=-\epsilon c_{\psi_3}=c_{\text{static}}\,.
\ee
Thus, $c_{\hat{\chi}}$ exactly matches $c_{\text{static}}$ \eqref{central-charge-static}.

This matching supports the claim that the chiral CFT appearing in Kerr/CFT is the DLCQ of the one appearing in the EVH/CFT \cite{Balasubramanian:2009bg}. Moreover, the Kerr-CFT also matches the temperature of the left-sector $T_L$ of the 2d CFT in EVH/CFT satisfying $\pi\ell_3 T_L=x_0/\ell_3$. (Note that in the extremal case the temperature of the right-moving sector of the 2d CFT appearing in EVH/CFT vanishes.) One may then use Cardy's formula
$$
S=\frac{\pi^2}{3} c (\ell_3 T_L)=\frac{\pi}{3}\cdot \frac{3q^2}{\ell^4\sqrt{\mathbf{V}_s}} N^2\epsilon\cdot \frac{x_0}{\ell_3}=\pi \frac{q}{\ell^2} \frac{\hat r_0}{\ell} N^2\epsilon=S_{\text{BH}}\,,
$$
where in the last equality we used \eqref{entropy-temp-static-like}.

Unlike the rotating case,  $c_\phi$ and $c_\psi$ also remain finite in the near-EVH static limit. They satisfy the relations $c_\phi=c_{\hat{\chi}}\cdot \ell \hb/q$, $c_\psi=c_{\hat{\chi}}\cdot \ell \ha/q$. Notice the proportionality coefficients agree with those appearing in the coordinate transformation \eqref{phi-psi-NEVH-static-scaling} removing the mixing between the angles on \sph{3} and the \ads{3} coordinates $\tau,\ \tilde{\chi}$. Within the Kerr/CFT mentality, one may then propose that in the near horizon, near-EVH static case we have three chiral CFT descriptions, one associated with the EVH/CFT via the DLCQ description and the other two (related to $c_\phi$ and $c_\psi$) with rotations on the \sph{3}. This latter, if true, may not be argued for through the standard Kerr/CFT prescription for computing
the central charges, which involves imposing certain boundary conditions for metric fluctuations
\cite{Kerr-CFT,Kerr-CFT-review}. To see this we note that the extremal black hole geometry we discuss here is
extremal-BTZ$\times$\sph{3}, the near horizon limit of which is \ads{3}-selfdual-orbifold$\times$\sph{3}.
This suggests that one should be able to extend the standard Kerr/CFT prescription to compute
the central charge to the cases like extremal-BH$\times X$, where $X$ is a compact geometry. This of course cries for
a thorough study and better understanding which we postpone to future works.

\section{Discussion}
\label{discussion-section}

To understand better the physics of Extremal Vanishing Horizon (EVH) black holes, and in particular the EVH/CFT proposal \cite{KKEVH}, we continued the analysis of \cite{static case} and extended it to stationary black holes in the class of asymptotic \ads{5} black holes. These are black holes with two equal electric charges and two independent spins. We classified all EVH black holes in this class and argued that generically the EVH hypersurface is a bifurcate co-dimension two surface. The bifurcation line, which corresponds to the case with vanishing spin, the static EVH black hole, is then a co-dimension three surface. We studied excitations around any given EVH point and showed that all these excitations can be captured by the near-horizon geometry, which has a locally \ads{3} throat, a pinching \ads{3} \cite{Massless-BTZ}.

We showed EVH black holes interpolate between \ads{5} at the boundary and a (locally) \ads{3} throat at the horizon and discussed the connection between the UV ${\cal N}=4$ $\U(N)$ SYM and the IR \cft{2} appearing in the EVH/CFT proposal.  Based on the arguments and proposal made in \cite{Massless-BTZ} we argued that to resolve the pinching orbifold we should take a large $N$ limit in the dual gauge theory such that both the temperature of the black hole, measured in 5d (or 10d) Planck units, and its entropy, remain finite. It is still desirable to have a better understanding of  the pinching resolution proposal made in \cite{Massless-BTZ}.

Although we did not fully specify the IR CFT${}_2$, we mentioned that it can be understood
as a specific BMN-type sector in the UV CFT$_4$ in the specific large N limit \eqref{N-scaling}. This
proposal, already made in  \cite{Fareghbal:2008ar,static case} for different sectors, should still be established through
explicit computations. In the limit in which the angular momentum is much larger than the R-charge, i.e. $a,b\to \ell$, evidence for the existence of a chiral spectrum of excitations was provided in  \cite{Berkooz:2012qh}.

As pieces of evidence for the EVH/CFT proposal we showed that the first law of thermodynamics for the original 5d (or 10d) EVH black hole, in the near-EVH limit
reduces to the first law of thermodynamics of the BTZ black hole appearing in the near horizon. This result is remarkable, not only because of non-trivial cancellations which happen at a technical level, but also because it holds quite generically  regardless of the details of the EVH black hole geometry \cite{to-appear}.

We also discussed a connection between the EVH/CFT proposal and Kerr/CFT for extremal excitations of EVH black holes, i.e. extremal near-EVH black holes.
We showed explicitly that the chiral CFT appearing in the Kerr/CFT proposal for extremal near-EVH black holes can
be understood as the DLCQ of the \cft{2} appearing in the EVH/CFT correspondence, realizing the proposal made in
\cite{Balasubramanian:2009bg}. There are several questions and points which asks for further analysis.
One closely related idea, providing a ``microscopic description''
for Kerr/CFT through locally \ads{3} throats, has also appeared in \cite{Kerr-CFT-review,microscopic-Kerr-CFT}.

There are two further points concerning this work that require further study : the identification of IR charges for the static EVH black holes and its connection with Kerr/CFT.

Regarding the identification of IR charges, it is known that the appearance of constant electric and magnetic fields which shift the values for the stress tensor and $\U(1)$ R-symmetry currents under spectral flow in the standard D1-D5 dual CFT \cite{Hansen:2006wu}. We suspect the same, if not more general set of spectral flows, should occur here accounting for the extra energy and angular momentum contributions in \eqref{J-chi-static-like} and \eqref{Delta-IR-static-like}. To understand this point, one must study the reduction of our 10d near horizon geometries to three dimensions, extending the reduction to six dimensions done in \cite{Fareghbal:2008ar}. As the 1st law must be invariant under these flows, the derivation of the IR first law in \ref{sec:staticf} should not be modified by spectral flow.

Regarding the connection to Kerr/CFT,  one can show that taking a near horizon of a given extremal black hole and taking an EVH limit do not commute. That is, there seems to be more than one geometry described by the same ``dual CFT'' (within the Kerr/CFT proposal). A similar feature has been reported in
the ``subtracted geometry'' proposal \cite{subtracted-geometry}, that one may ``deform'' the near horizon geometry
without changing the Kerr/CFT description. It is desirable to study a possible connection between the ideas discussed in these papers and the one we presented here.

\section*{Acknowledgements}
MJ thanks the KIAS in Seoul for hospitality during the completion of this work. MJ and JS would like to thank the KITP in Santa Barbara for hospitality during the completion of this work. JS would like to thank Nordita, WITS and the University of Cape Town for hospitality during the final stages of this work. MMSHJ would like to thank WITS university for hospitality at the early stages of this work. MMSHJ, JS and HY would like to thank CQUeST and the \emph{Quantum Aspects of Black Holes} workshop in Seoul for the hospitality in the last stages of this work. The work of MJ and JS was partially supported by the Engineering and Physical Sciences Research Council (EPSRC) [grant number EP/G007985/1] and the Science and Technology Facilities Council (STFC) [grant number ST/J000329/1]. The  work of HY was supported by the National Research Foundation of Korea Grant funded by the Korean Government (NRF-2011- 0023230).

\appendix

%%%%%%%%%%%%%%
\section{Horizon Structure}
\label{ap:horizon}
%%%%%%%%%%%%%%

Whenever the equation $X(r)=0$ allows real solutions, the configurations \eqref{5d BH metric} describe a family of black holes. When this is not the case, it describes a naked singularity.
In this appendix, we study the constrains in parameter space for black holes to exist. To do so, define
\bea
\label{Xr}
 && l^2  X \equiv  r^4+Ar^2+B+\frac{C}{r^2}  \\
 && = r^4 + \left[\ell^2+a^2+b^2+2q \right]r^2+\left[ (a^2+b^2)(\ell^2+q)+a^2b^2+q^2-2m\ell^2  \right]+\frac{a^2b^2\ell^2}{r^2}\nonumber
 \eea
Note $A, C\in \RR^+$, because $a,b\in \RR$ and $q\in\RR^+$, whereas $B$ can be negative for large $m$.

We shall denote the outer and inner horizons by $r_\pm$. These correspond to the largest and smallest positive roots of the equation $X(r)=0$. When $r_+=r_-$,  \eqref{5d BH metric} corresponds to extremal black holes. Figure \ref{X} shows the root structure for the equation $X(r)=0$. The existence of a horizon requires $X_c \leq 0$, where $X_c$ is the extremum of $X$. This constraints the parameters $a, b, q$ and $m$.

Charges carried by the EVH black holes studied in this work are such that $C \ll |B|\ll A$. When these hold, $r_\pm$
can be expanded as follows
\be
r_\pm^2=-\frac{B}{2A}\pm \sqrt{\left(\frac{B}{2A}\right)^2-\frac{C}{A}} + \cdots .
\ee
Existence of horizons requires $B^2 \geq 4AC$.
 \begin{figure}[h]
 \begin{center}
  \includegraphics[scale=1]{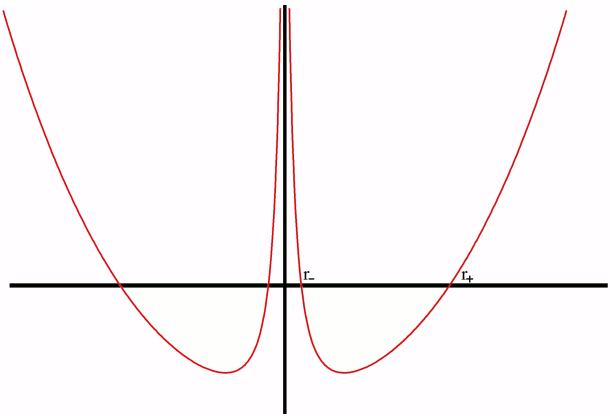}
 \caption{$X$ as a function $r$ for a general non-extremal black hole solution. }
 \label{X}
 \end{center}
 \end{figure}

Near-EVH solutions occur when $r_\pm\sim \epsilon\to 0$. This can happen if (note that $A$ is finite)
\begin{align}
\mathrm{Rotating EVH}:&\quad   a =\text{finite} ,\;\; {b} = \hb\epsilon^{2\alpha}, \;\; B=-\hat{B} \epsilon^{2}\,,\qquad \alpha\geq 1 \\
\mathrm{Static EVH}:&\quad   {a} = \ha\epsilon^{\beta},\;\; b=\hb\epsilon^{\alpha}, \;\; B=-\hat{B}\epsilon^{2}\,,\qquad
\alpha\geq\beta\geq 1,
\end{align}
where $\hat{B}\geq 0$ and $C=\hat{C}\epsilon^{4+\gamma},\ \gamma\geq 0$. These are of course the two cases
discussed in (\ref{EVHgen}) and (\ref{HX2}).

\paragraph{Rotating EVH:}
To study this case let us assume $\alpha=1,\ \gamma=0$. Indeed it is not difficult to show that  $\alpha>1$ cases can be recovered from $\alpha=1$, by sending $\hb\to 0$ and we get  back to solution  (\ref{NHEVHKK}).
From (\ref{Xr}) we can read parameter $m$
\be
\label{m}
m=\frac{1}{2\ell^2} (\ell^2a^2+qa^2+q^2)+\frac{\hat{B}}{2\ell^2}\epsilon^2 + \frac{\hb^2(\ell^2+a^2+q)}{2l^2}\epsilon^4.
\ee
For the above parameters, condition $B^2\geq 4AC$ is given by
\be
\hat{B}^2 \geq 4a^2\hb^2\ell^2(\ell^2+2q+a^2)+{\mathcal O}(\epsilon^2)
\ee
\paragraph{Static EVH:}
For this case, without loss of generality we can assume $\alpha=\beta=1,\ \gamma=0$ (larger values of $\alpha,\beta$ and $\gamma$ may be
obtained from this case in the $\ha,\hb,\ \hat{C}\to 0 $ limit).
In this case the $B^2\geq 4AC$ condition reduces to
\be
\hat{B}^2\geq (\ell^2+2q)\ha^2\hb^2\ell^2\,,
\ee
and the negative $\hat{B}$ condition implies,
\be
2m\ell^2-q^2\geq (\ell^2+q)(\ha^2+\hb^2)\epsilon^2\,.
\ee

Finally, we note that for generic values of the parameters, the black hole horizon topology
is \sph{3} or \sph{3}$\times$\sph{5}, depending on whether one takes the 5d or 10d perspective.
For the specific values discussed in section \ref{sec:EVH}, they degenerate to \sph{2}$\times$\sph{1}.  Furthermore,
our black hole configurations \eqref{5d BH metric} have closed time-like curves.
For more discussions on these black holes and their singularity and causal structure, see
\cite{Chong:2005da,More-on-AdS5-bh,Chow:2008dp}.

%%%%%%%%%%%%%%%%%%%%%%%%%%%%%%%%
\section{Near Horizon geometries as 5 dimensional geometries}
\label{Appendix-5d-NH-limit}
%%%%%%%%%%%%%%%%%%%%%%%%%%%%%%%%

As discussed in section \ref{section:NH-KK-EVH}, for the rotating EVH case there is a freedom in choosing
the $\chi$ angle. In particular, one may choose it to be $\psi$, corresponding to $\omega_\chi=\pi/2$ (see Fig. \ref{angles}). With this choice, the near horizon geometry may be taken over the 5d black hole solution \eqref{5d BH metric} without considering the 10d uplift. To this end, consider the following scalings
\bea
r=\frac{a}{\sqrt{a^2+q}}\epsilon x, \quad t= \frac{\sqrt{a^2+q}}{a}\frac{\tau}{\epsilon},\quad \psi=\frac{\tilde{\psi}}{\epsilon},\quad \phi=\tilde{\phi}+\frac{l^2-a^2}{l^2 a \sqrt{a^2+q}}\frac{\tau}{\epsilon}
\eea
Taking $\epsilon\rightarrow 0$, we get following geometry
\bea
&& ds^2=\frac{a^{2\over 3}\cos^{2\over 3}\theta}{(q+a^2)^{1\over 3}}h_1^{\frac{4}{3}} \bigg[-\frac{(x^2-x_+^2)(x^2-x_-^2)}{\ell_3^2x^2}d\tau^2+\frac{\ell_3^2x^2dx^2}{(x^2-x_+^2)(x^2-x_-^2)}+x^2(d\tilde{\psi}-\frac{x_+x_-}{\ell_3x^2}d\tau)^2\bigg] \nonumber \\
&&\quad \frac{a^{2\over 3}\cos^{2\over 3}\theta}{\Delta_{\theta}}(a^2+q)^{2\over 3}h_1^{4\over 3}\left( d\theta^2 + \frac{\Delta_{\theta}^2\sin^2\theta}{\Sigma_a^2h_1^4}d\tilde{\phi}^2 \right)
\eea
Scalar fields and non-zero components of gauge fields in this limit are given by
\bea
&& X_1=X_2=\frac{a^{2\over 3}\cos^{2\over 3}\theta}{(a^2+q)^{1\over 3}h_1^{2\over 3}}, \quad X_3=\frac{(a^2+q)^{2\over 3}h_1^{3\over 3}}{a^{4\over 3}\cos^{4\over 3}\theta}\\
&& F_{\theta\tilde{\phi}}^{(1)}=F_{\theta\tilde{\phi}}^{(2)}=-\frac{2a\sqrt{q(1+\frac{q}{l^2})}\sin\theta\cos\theta}{\Sigma_a h_1^4\sqrt{q+a^2}}.
\eea

%%%%%%%%%%%%%  BIBLIOGRAPHY  %%%%%%%%%%%%%

\end{document}